\documentclass{sig-alternate-10pt}

\usepackage{url}
\usepackage{mathrsfs}
\usepackage{enumitem}
\usepackage{MnSymbol}
\usepackage{amsbsy}
\usepackage{clrscode}

\newcounter{descriptcount}

\newtheorem{definition}{Definition}
\newtheorem{theorem}{Theorem}
\newtheorem{lemma}{Lemma}
\newtheorem{proposition}{Proposition}

\newtheorem{corollary}{Corollary}

\def\core{\mathop{\mathit core}}
\def\dom{\mathop{\mathit dom}}
\def\chase{\mathop{\mathit Chase}}

\def\word{\mathop{\mathit word}}
\def\paths{\mathop{\mathit paths}}

\def\arity{\mathop{\mathit arity}}

\newcommand{\sctaa}{\ensuremath{{{\mathsf{CT}}^{\mathsf{std}}_{\forall\forall}}}}
\newcommand{\sctae}{\ensuremath{{{\mathsf{CT}}^{\mathsf{std}}_{\forall\exists}}}}
\newcommand{\sctia}[1]{\ensuremath{{{\mathsf {CT}}^{\mathsf{std}}_{{#1},\forall}}}}
\newcommand{\sctie}[1]{\ensuremath{{{\mathsf {CT}}^{\mathsf{std}}_{{#1},\exists}}}}

\newcommand{\octaa}{\ensuremath{{{\mathsf{CT}}^{\mathsf{obl}}_{\forall\forall}}}}
\newcommand{\octae}{\ensuremath{{{\mathsf{CT}}^{\mathsf{obl}}_{\forall\exists}}}}
\newcommand{\octia}[1]{\ensuremath{{{\mathsf {CT}}^{\mathsf{obl}}_{{#1},\forall}}}}
\newcommand{\octie}[1]{\ensuremath{{{\mathsf {CT}}^{\mathsf{obl}}_{{#1},\exists}}}}

\newcommand{\soctaa}{\ensuremath{{{\mathsf {CT}}^{\mathsf{sobl}}_{\forall\forall}}}}
\newcommand{\soctae}{\ensuremath{{{\mathsf {CT}}^{\mathsf{sobl}}_{\forall\exists}}}}
\newcommand{\soctia}[1]{\ensuremath{{{\mathsf {CT}}^{\mathsf{sobl}}_{{#1},\forall}}}}
\newcommand{\soctie}[1]{\ensuremath{{{\mathsf {CT}}^{\mathsf{sobl}}_{{#1},\exists}}}}

\newcommand{\cctaa}{\ensuremath{{{\mathsf {CT}}^{\mathsf{core}}_{\forall\forall}}}}
\newcommand{\cctae}{\ensuremath{{{\mathsf {CT}}^{\mathsf{core}}_{\forall\exists}}}}
\newcommand{\cctia}[1]{\ensuremath{{{\mathsf {CT}}^{\mathsf{core}}_{{#1},\forall}}}}
\newcommand{\cctie}[1]{\ensuremath{{{\mathsf {CT}}^{\mathsf{core}}_{{#1},\exists}}}}

\newcommand{\actaa}{\ensuremath{{{\mathsf{CT}}^{\star}_{\forall\forall}}}}
\newcommand{\actae}{\ensuremath{{{\mathsf{CT}}^{\star}_{\forall\exists}}}}
\newcommand{\actia}[1]{\ensuremath{{{\mathsf {CT}}^{\star}_{{#1},\forall}}}}
\newcommand{\actie}[1]{\ensuremath{{{\mathsf {CT}}^{\star}_{{#1},\exists}}}}

\newcommand{\depenrich}[1]{\ensuremath{\hat{#1}}}
\newcommand{\enrich}[1]{\ensuremath{\widehat{#1}}}
\newcounter{foo}
\newcounter{qcounter}

\begin{document}

\title{Anatomy of the chase}

\numberofauthors{2} 
\author{
\alignauthor G\"{o}sta Grahne\\
\affaddr{Concordia University}\\
\affaddr{Montreal, Canada, H3G 1M8}\\
\email{grahne@cs.concordia.ca}
\alignauthor Adrian Onet
\titlenote{Contact author.}\\
\affaddr{Concordia University}\\
\affaddr{Montreal, Canada, H3G 1M8}\\
\email{a\_onet@cs.concordia.ca}
}

\maketitle

\begin{abstract}

\bigskip
A lot of research activity has recently taken place around the chase procedure, 
due to its usefulness in data integration, data exchange, query optimization, 
peer data exchange and data correspondence, to mention a few. As the chase 
has been investigated and further developed by a number of research groups 
and authors, many variants of the chase have emerged and associated results 
obtained. Due to the heterogeneous nature of the area it is frequently 
difficult to verify the scope of each result. In this paper we take closer 
look at recent developments, and provide additional results.
Our analysis allows us create 
a taxonomy of the chase variations and the properties they satisfy. 

Two of the most central problems regarding the chase is termination, 
and discovery of restricted classes of sets of dependencies that guarantee 
termination of the chase. 
The search for the restricted classes has been motivated by 
a fairly recent result that shows that it is undecidable to 
determine whether the chase with a given dependency set will 
terminate on a given instance. 
There is a small dissonance here, since the quest has been 
for classes of sets of dependencies guaranteeing termination of the chase 
on {\em all} instances, even though the latter problem was not 
known to be undecidable. 
We resolve the dissonance in this paper by showing that determining 
whether the chase with a given set of dependencies terminates on all 
instances 
is {\sf coRE}-complete. Our reduction also gives us the aforementioned 
instance-dependent {\sf RE}-completeness result as a byproduct. 
For one of the restricted classes,
the {\em stratified} sets dependencies,
we provide new complexity results for
the problem of testing whether a given set
of dependencies belongs to it.
These results rectify some previous claims that have occurred
in the literature.
\end{abstract}

\bigskip


\keywords{Chase, Date Exchange, Data Repair, Incomplete databases, Undecidability Complexity} 


\bigskip
\bigskip
\section{Introduction}\label{introduction}

\bigskip
The chase procedure was initially developed for testing
logical implication between sets of dependencies  
\cite{DBLP:conf/icalp/BeeriV81}, for determining 
equivalence of database instances known to satisfy 
a given set of dependencies 
\cite{DBLP:conf/xp/Mendelzon81,DBLP:journals/jacm/ImielinskiL84},
and for determining query equivalence under database 
constraints \cite{320091}.
Recently the chase has experienced a revival due to
its application in data integration, data exchange, data 
repair, query optimization, ontologies and data correspondence.
In this paper we will focus on 
constraints in the form of
embedded dependencies
\cite{DBLP:journals/jacm/Fagin82}
specified by sets of tuple generating dependencies (tgd's).
A tgd is a first order formula of the form
$$
\forall \bar{x},\bar{y}\; \big( \alpha(\bar{x},\bar{y}) 
\rightarrow \exists \bar{z}\; \beta(\bar{x},\bar{z}) \big),
$$
where $\alpha$ and $\beta$ are conjunctions of relational 
atoms, and $\bar{x}$, $\bar{y}$, and $\bar{z}$ are sequences of variables.
We refer to $\alpha$ as the body and $\beta$ as the head of the dependency.
Sometimes, for simplicity, the tgd is written as  $\alpha\rightarrow \beta$.
Intuitively the chase procedure repeatedly applies a series of 
chase steps to database instances that violate
some dependency. 
Each such chase step takes a tgd that is not satisfied by the 
instance, a set of tuples that witness the violation,
and adds new tuples
to the database instance so that the resulting instance 
does satisfy the tgd with respect to those witnessing tuples. 

Given an instance $I$ and a set of tgd's $\Sigma$, 
a {\em model} of $I$ and $\Sigma$ is a database instance
$J$ such that there is a homomorphism from $I$ to $J$, 
and $J$ satisfies $\Sigma$.
A {\em universal model} of $I$ and $\Sigma$ is
a {\em finite} model of $I$ and $\Sigma$
that has a homomorphism into {\em every}
model of $I$ and $\Sigma$.
It was shown in \cite{DBLP:conf/icdt/FaginKMP03,DBLP:conf/pods/DeutschNR08} 
that the chase computes a universal model
of $I$ and $\Sigma$, whenever $I$ and $\Sigma$
has one. In case
$I$ and $\Sigma$ does not have a universal model
the chase doesn't terminate (in this case 
it actually converges at a countably infinite model).


As the research on the chase has progressed
several variations of the chase have evolved.
As a consequence it has become difficult
to determine the scope of the results obtained.
We scrutinize the four most important chase variations,
both deterministic and non-deterministic. 
We will check for each of these chase variations
the data and combine complexity of testing
if the chase step is applicable 
for a given instance and tgd.
It didn't came as a surprise to
find out that the 
oblivious and semi-oblivious chase
variation share the same complexity,
but the standard chase has a slightly 
higher complexity.
The table below shows the data and combined complexity
for the following problem:
given an instance with $n$ tuples
and a tgd ($\alpha\rightarrow \beta$), 
is the standard/oblivious chase step applicable?

\medskip 
\begin{center}
\begin{tabular}{lll} 
 & {\ \ \ \ Data} & {\ Combined}  \\
Chase & {Complexity} & {Complexity}  \\
\hline
\hline
standard & $O(n^{|\alpha|+|\beta|})$ & $\Sigma^P_2$-complete  \\
oblivious & $O(n^{|\alpha|})$ & {\sf NP}-complete  \\
\hline     
\end{tabular}
\end{center}

\medskip 
Thus, at a first look the oblivious and semi-oblivious 
chase procedures will be a more appropriate 
choice when it comes to a practical implementation.
Still, as we will show,
 the lower complexity comes with 
a price, that is  
the higher the complexity
for a chase variation
the more "likely" is the chase process 
to terminate for a given instance and set of dependencies.
Thus, the core chase,
that not only applies in parallel all standard
chase steps but it also computes the core of the 
resulted instance, has the highest complexity
of the chase step.
On the other hand 
from \cite{DBLP:conf/pods/DeutschNR08}
we know that 
the core chase is complete in finding 
universal models meaning that
if any of the chase variations
terminates for some input,
then the core chase terminates as well.
We next compare the semi-oblivious and
standard chase when it comes to the termination problem.
With this we show 
that the standard and semi-oblivious chase 
are not distinguishable
for the most classes of dependencies developed
to ensure the standard chase termination.
Furthermore,
we show that 
the number of semi-oblivious chase steps
needed to terminate  remains 
the same as for the standard chase,
namely polynomial.
This raises the following question: 
What makes a class of dependency sets
that terminate for all input instances under 
the standard chase to terminate for the
semi-oblivious chase as well?
We answer this question by giving a 
sufficient syntactical condition 
for  classes of dependency sets
that ensures termination on all instances for the standard 
chase to also guarantee termination for the semi-oblivious chase.
As we will see most of the  known
classes of dependencies build to ensure the standard chase 
termination on all instances are actually
guaranteeing termination for the 
less expensive semi-oblivious chase variation.

It has been known for some time
\cite{DBLP:conf/pods/DeutschNR08,DBLP:conf/kr/CaliGK08,DBLP:conf/pods/Marnette09}
that it is undecidable to determine if the chase with a given set
of tgd's terminates on a given instance. 
This has spurred a quest for restricted classes of
tgd's guaranteeing termination.
Interestingly, these classes all guarantee {\em uniform} termination,
that termination on {\em all} instances. 
This, even though it was 
only known that the problem is undecidable for
a given instance. We remediate the situation
by proving that 
(perhaps not too surprisingly)
the uniform version of the termination problem is
undecidable as well, and show that it is not
recursively enumerable. 
We show that the determining whether
the core chase with a given set of dependencies
terminates on all instances 
is {\sf coRE}-complete.
We achieve this using a reduction
from the uniform termination problem
for word-rewriting systems (semi-Thue systems).
As a byproduct we obtain the result from 
\cite{DBLP:conf/pods/DeutschNR08} showing that testing
if the core chase terminates for a given instance and 
a given set of dependencies is {\sf RE}-complete.
We will show also that the same complexity result
holds for testing whether the standard chase
with a set of dependencies
is terminating on at least one
execution branch.
Next we will show that by using a single denial constraint
(a ``headless'' tgd)
in our reduction the same complexity result holds also for 
the standard chase termination on all instances on all execution 
branches. 
It remains an open problem if this holds without
denial constraints.

Many of the restricted classes guaranteeing termination
rely on the notion of a set $\Sigma$ of dependencies
being {\em stratified}. Stratification involves
two conditions, one 
determining a partial order between tgd's in
in $\Sigma$,
and the other on $\Sigma$ as a whole. 
It has been claimed that testing 
the partial order between tgd's 
is in {\sf NP}
\cite{DBLP:conf/pods/DeutschNR08}.
We show that this cannot be the case
(unless {\sf NP}={\sf coNP}), by proving
that the problem is at least {\sf coNP}-hard.
We also prove a $\Delta^p_2$
upper bound
for the problem.
Finding matching upper and lower bounds
remains an open problem.

\subsubsection*{Paper outline}

\smallskip

The next section contains the preliminaries
and describes the chase procedure and its variation.
Section 3 deals with the 
complexity of testing if for an instance and a dependency
there exists an ``applicable'' chase step.
Section 4 deals with problems related to the chase termination. 
We define termination classes for each of
the chase variations and then determine the
relationship between these classes.
Section 4 also contains our main result, namely,
that it is {\sf coRE}-complete to test if the chase 
variations with a given 
set of dependencies 
terminate on all instances. 
This result is obtained via a 
reduction from the uniform termination problem for
word-rewriting systems.
In Section 5 we review the main
restricted classes that ensure termination
on all instances,
and relate them to different chase variations
and their termination classes.
Finally, in Section 6
we provide complexity results related to 
the membership problem for the 
stratification based classes of dependencies 
that ensure the standard chase termination.
Conclusions and further work are drawn in the last section.
Proofs not given in the paper 
are included in an Appendix.

\bigskip
\bigskip
\section{Preliminaries}\label{prelim}

\bigskip
For basic definitions and concepts we
refer to \cite{DBLP:books/aw/AbiteboulHV95}. We will consider
the complexity classes
{\sf PTIME},
{\sf NP},
{\sf coNP},
{\sf DP},
{\sf RE},
{\sf coRE},
and the first few levels
of the polynomial hierarchy.
For the definitions of these classes
we refer to \cite{DBLP:books/daglib/0072413}.
 
We start with some preliminary notions.
We will use the symbol $\subseteq$ for 
the subset relation, and $\subset$
for proper subset.
A function $f$ with a finite set
$\{x_1,\ldots,x_n\}$ as domain,
and where $f(x_i)=a_i$,
will be described as
$\{x_1/a_1,\ldots,x_n/a_n\}$. 
The reader is cautioned that the symbol
$\rightarrow$ will be overloaded;
the meaning should however be clear
from the context. 

\noindent
{\bf Relational schemas and instances}.
A {\em relational schema} is a finite set
$\mathbf{R} = \{R_1,\ldots,R_n\}$ of relational symbols
$R_i$, each with an associated positive integer $\arity(R_i)$.
Let {\sf Cons} be a countably infinite set of
constants, usually denoted $a,b,c,\ldots$,
possibly subscripted, and let {\sf Nulls} be
a countably infinite set of nulls denoted
$x,y_1,y_2,\ldots$. 
A {\em relational instance}
over a schema $\mathbf{R}$ is a function that
associates for each relational symbol $R\in\mathbf{R}$
a finite subset $R^{I}$ of 
$({\sf Cons} \; \cup \; {\sf Nulls})^{\arity(R)}$.

For notational convenience we shall 
frequently identify 
an instance $I$ with the
set $\{R(\bar{a})  : (\bar{a})\in R^I, R\in\mathbf{R}\}$
of atoms, assuming appropriate lengths 
of the sequence $\bar{a}$ for each
$R~\in~\mathbf{R}$.
By the same convenience,
the atoms $R(a_1\ldots,a_k)$ will be called {\em tuples}
of relation $R^I$ and denoted $t,t_1,t_2,\ldots$.
By $\dom(I)$ we mean the set of
all constants and nulls occurring in the
instance $I$, and by $|I|$ we mean the number
of tuples in $I$. 

\bigskip
\noindent
{\bf Homomorphisms}.
Let $I$ and $J$ be instances,
and $h$ a mapping from $\dom(I)$ to $\dom(J)$
that is the identity on the constants.
We extend $h$ to tuples $(\bar{a})=(a_1,\ldots,a_k)$ by
$h(a_1,\ldots,a_k) = (h(a_1),\ldots,h(a_k))$.
By our notational convenience we can thus
write $h(\bar{a})$ as $h(R(\bar a))$,
when $(\bar{a})\in R^I$.
We extend $h$ to instances by
$h(I) = \{h(t) : t\in I\}$.
If $h(I)\subseteq J$ we say that 
$h$ is a {\em homomorphism} from $I$ to $J$.
If $h(I)\subseteq I$,
we say that $h$
is an {\em endomorphism}, and if $h$ also 
is idempotent, $h$ is called a {\em retraction}.
If $h(I)\subseteq J$,
and the mapping $h$ is a bijection,
and if also $h^{-1}(J)~=~I$,
the two instances
are {\em isomorphic}.

A subset $I'$ of $I$ is said to be a
{\em core} of $I$, if there is a
endomorphism $h$, such that $h(I)\subseteq I'$,
and there is no
endomorphism $g$ such that
$g(I')\subset I'$.
It is well known that all cores
of an instance $I$ are isomorphic,
so for our purposes we can consider 
the core unique,
and denote it $\core(I)$.

\bigskip
\noindent
{\bf Tuple generating dependencies}.
A {\em tuple generating dependency  {\em (}tgd{\em )}} is
a first order formula of the form
$$
\forall \bar{x},\bar{y}\; 
\big( \alpha(\bar{x},\bar{y})\rightarrow 
\exists \bar{z}\;
\beta(\bar{x},\bar{z}) \big),
$$
where $\alpha$ and $\beta$ are 
conjunctions of relational atoms,
and $\bar{x}$,$\bar{y}$ and $\bar{z}$ 
are sequences of variables.
We assume that
the variables occurring in tgd's come from a countably
infinite set {\sf Vars} disjoint from {\sf Nulls}.
We also allow constants in the tgd's.
In the formula we call $\alpha$ the {\em body} of the tgd.
Similarly we refer to $\beta$ as the {\em head}
of the~tgd. If there are no existentially
quantified variables the dependency is said
to be {\em full}.

When $\alpha$ is the body of a tgd
and $h$ a mapping from the set
${\sf Vars}\hspace{0.03cm}\cup\hspace{0.03cm}{\sf Const}$
to ${\sf Nulls}\hspace{0.03cm}\cup\hspace{0.03cm}{\sf Const}$
that is identity on constants, 
we shall conveniently regard 
the set of atoms in $\alpha$ as an instance $I_{\alpha}$,
and write $h(\alpha)$ for 
the set $h(I_{\alpha})$. 
Then $h$ is a homomorphism
from $\alpha$ to an instance $I$,
if $h(\alpha)\subseteq I$.


Frequently, we omit the universal quantifiers
in tgd formulas.
Also, when the variables and constants 
are not relevant
in the context, we 
denote a tuple generating dependency 
$
\alpha(\bar{x},\bar{y})\rightarrow 
\exists \bar{z}\;
\beta(\bar{x},\bar{z})
$
simply as $\alpha\!\rightarrow\!\beta$. 

Let $\xi = \alpha\!\rightarrow\!\beta$  
be a tuple generating dependency,
and $I$ be an instance. Then we say that
$I$ {\em satisfies} $\xi$, if 
$I\models\xi$
in the standard model theoretic sense, 
or equivalently, 
if for every homomorphism $h$, such that
$h(\alpha)\subseteq I$, there is an extension
$h'$ of $h$, such that $h'(\beta)\subseteq I$.


\bigskip
\noindent
{\bf The Chase}.
Let $\Sigma$ be a (finite) 
set of tgd's and $I$ an instance. 
A~{\em trigger} for the set $\Sigma$ on $I$ 
is a pair $(\xi,h)$, where 
$\xi = $
$\alpha\!\rightarrow\!\beta\in\Sigma$,
and $h$ is a homomorphism such that
$h(\alpha)\subseteq I$.
If, in addition, 
there is no extension $h'$ of $h$,
such that $h'(\beta) \subseteq I$,
we say that the trigger
$(\xi,h)$ is {\em active} on $I$.

Let $(\xi,h)$ be a trigger for $\Sigma$ on $I$. 
To {\em fire} the trigger means
transforming $I$ into
the instance 
$J=I\cup\{h'(\beta)\}$, where $h'$ is 
a {\em distinct extension} of $h$,
i.e.\ an extension
of $h$ that assigns new fresh nulls
to the existential variables in $\beta$.
By ``new fresh'' we mean the next unused element
in some fixed enumeration
of the nulls.
We denote this transformation 
$I\xrightarrow{(\xi,h')}J$,
or just 
$I\rightarrow J$,
if the particular trigger is 
irrelevant or understood from the context.

A sequence $I_0,I_1,I_2\ldots$ of instances (finite or infinite)
is said to be a {\em chase sequence with
$\Sigma$ originating from} $I_0$, if
$I_i\rightarrow I_{i+1}$  for all $i=0,1,2,\ldots$.
At each step there can naturally be several triggers to
choose from, so in general there will be several
chase sequences originating from $I_0$ for
any given set $\Sigma$ of tgd's. 
If for some $i$ that is there are no
more triggers to be fired for $I_i$, 
we say that the sequence {\em terminates}.
Otherwise the sequence is infinite.

In summary, the chase process can be seen as a tree
rooted at $I_0$, and with the individual chase 
sequences as branches.
From an algorithmic point of view the choice
of the next trigger to fire
is essential.
Based on this, the following variations
of the chase process have been 
considered in the literature. 
(for a comprehensive review of different chase variation see \cite{Onet12}).

\medskip 
\begin{enumerate}
\item
{\em The standard chase} \cite{DBLP:conf/icdt/FaginKMP03}.
The next trigger is chosen nondeterministically
from the subset of current triggers that are active.

\medskip
\item
{\em The oblivious chase} \cite{DBLP:conf/kr/CaliGK08}.
The next trigger is chosen nondeterministically
from the set of all current triggers, active or not,
but each trigger is fired only once in a chase sequence.

\medskip
\item
{\em The semi-oblivious chase} \cite{DBLP:conf/pods/Marnette09}.
Let $\xi$ be a tgd $\alpha(\bar{x},\bar{y})\rightarrow\beta(\bar{x},\bar{z})$.
Then triggers $(\xi,h)$ and $(\xi,g)$ are considered
equivalent if $h(\bar{x})=g(\bar{x})$.
The semi-oblivious chase works as the oblivious one,
except that exactly one trigger
from each equivalence class is fired in a branch.

\medskip
\item
{\em The core chase} \cite{DBLP:conf/pods/DeutschNR08}.
At each step, all currently active triggers are fired
in parallel, and then the core of the 
union of the resulting instances is
computed before the next step.
Note that this makes the chase process deterministic.
\end{enumerate}

The three first variations are all nondeterministic,
but differ in which triggers they fire.
Also we consider all chase procedures to be {\em fair},
meaning that in any infinite chase sequence,
if a trigger is applicable at some chase step $i$,
then there exists an integer $j\geq i$
such that the trigger is fired at step $j$.

To illustrate the
difference between these chase variations, 
consider dependency set
 $\Sigma=\{\xi\}$, where
$\xi$ is tgd 
$R(x,y) \rightarrow \exists z\; S(x,z)$, 
and instance:

\begin{center}
\begin{tabular}{l} 
 \ \ \ \ $I_0$ \\
\hline 
$R(a,b)$ \\
$R(a,c)$ \\
$S(a,d)$
\end{tabular}
\end{center}

There are two triggers for the set $\Sigma$ on instance $I_0$,
namely $({\xi},{\{ x/a, y/b \}})$ and
$({\xi},{\{x/a, y/c \}})$.
Since $I_0\models\xi$ neither of the
triggers is active, so the standard chase
will terminate at $I_0$.
In contrast, both the oblivious and semi-oblivious chase
will fire the first trigger, resulting in instance
$I_1 = I_0\cup\{S(a,z_1)\}$.
The semi-oblivious chase will terminate at this point,
while the oblivious chase will fire the second trigger,
and then terminate in 
$I_2 = I_1\cup\{S(a,z_2)\}$. The core chase in this case will 
terminate also with $I_0$.

\bigskip
\section{Complexity of the chase step}\label{SEC:complexityChaseStep}


Algorithmically, 
there are two problems to consider.
For knowing when to terminate the chase, 
we need to
determine whether for a given instance $I$ and tgd $\xi$
there exists a homomorphism $h$ such that
$(\xi,h)$ is trigger on $I$.
This pertains to the
oblivious and semi-oblivious variations.
The second problem pertains to the standard
and core chase: given an instance $I$
and a tgd $\xi$, is there a homomorphism $h$,
such that $(\xi,h)$ is an {\em active} trigger  
on $I$. We call these problems the
{\em trigger existence} problem, 
and the {\em active trigger existence} problem,
respectively.
The {\em data complexity} of these problems
considers $\xi$ fixed, and in the {\em combined complexity}
both $I$ and $\xi$ are part of the input.
The following theorem gives the 
combined and data complexities of the two problems.

\begin{theorem}
Let $\xi$ be a tgd and $I$ an instance.
Then
\begin{enumerate}
\item
For a fixed $\xi$,
testing whether there exists a trigger
or an active trigger on a given $I$
is polynomial.
\item 
Testing whether there exists a trigger
for a given $\xi$ on a given $I$
is {\sf NP}-complete.
\item 
Testing whether there exists an active trigger 
for a given $\xi$ and a given $I$
is $\Sigma^p_2$-complete. 
\end{enumerate}
\end{theorem}

{\bf Proof}: (Sketch)
The polynomial cases can be verified by 
checking all homomorphisms from the body of the dependency 
into the instance. For the active trigger problem we also need to consider, 
for each such homomorphism, if it has an extension that maps
the head of the dependency into the instance.
These tasks can be carried out in 
$O(n^{|\alpha|})$ and  $O(n^{|\alpha|+|\beta|})$ time, respectively.

It is easy to see that the trigger existence problem
is {\sf NP}-complete in combined complexity,
as the problem is equivalent to testing 
whether there exists a homomorphism between two instances (in our case
$\alpha$ and $I$); 
a problem known to be {\sf NP}-complete.

For the combined complexity of the active trigger existence problem,
we observe that it is in $\Sigma^P_2$,
since one may guess a homomorphism $h$
from $\alpha$ into $I$, 
and then use an {\sf NP} oracle to verify
that there is no extension $h'$ of $h$,
such that $h'(\beta)\subseteq I$.
For the lower bound we
will reduce the following problem 
to the active trigger existence problem \cite{Rutenberg:1986:CGG:22416.22478}.
Let $\phi(\bar{x},\bar{y})$ be a Boolean formula in 3CNF
over the variables in $\bar{x}$ and $\bar{y}$.
Is the formula
$$
\exists \bar{x}\; \neg \big(\exists \bar{y} \, \phi(\bar{x},\bar{y})\big)
$$
true?
The problem is a variation 
of the standard $\exists\forall$-QBF problem
\cite{DBLP:journals/tcs/Stockmeyer76}.

For the reduction,   
let $\phi$ be given.
We construct an instance $I_{\phi}$
and a tgd 
$\xi_{\phi}$. 
The instance $I_{\phi}$ is as follows:

\medskip

\begin{minipage}[b]{0.50\linewidth}
\centering
\begin{tabular}{lll} 
\multicolumn{3}{c}{$F$}  \\
\hline 
1&0&0  \\ 
0&1&0  \\ 
0&0&1  \\ 
1&1&0  \\ 
1&0&1  \\ 
0&1&1  \\ 
1&1&1  

\end{tabular}
\end{minipage}
\begin{minipage}[b]{0.50\linewidth}
\centering
\begin{tabular}{ll} 
\multicolumn{2}{c}{$N$}  \\
\hline 
0&1 \\
1&0 \\
\\
\\
\\
\\
\\
\end{tabular}
\end{minipage}

\medskip

The tgd $\xi_{\phi} = \alpha\!\rightarrow\!\beta$
is constructed as follows.
For each variable $x \in \bar{x}$ in 
$\phi(\bar{x},\bar{y})$,
the
body $\alpha$ will
contain the atom  $N(x,x')$ 
($x'$ is used to represent~$\neg x$).
The head $\beta$ is existentially quantified
over that set 
$\bigcup_{y\in{\bar{y}}}\{ y,y'\}$
of variables. 
For each conjunct $C$ of $\phi$, 
we place in $\beta$ an atom
$F(x,y,z)$, 
where $x,y$ and $z$ are the variables in $C$, 
with the convention that if variable $x$ is
negated in $C$,
then $x'$ is used in the atom.
Finally for each $y \in \bar{y}$, 
we place in $\beta$ the atom $N(y,y')$,
denoting that $y$ and $y'$ 
should not have the same truth assignment.

Let us now suppose that the formula
$\exists \bar{x}\; \neg \big(\exists \bar{y} \, \phi(\bar{x},\bar{y})\big)$
is true. This means that there is a $\{0,1\}$-valuation $h$
of $\bar{x}$ such that for any $\{0,1\}$ valuation $h'$ of $\bar{y}$,
the formula
$\phi(h(\bar{x}),h'(\bar{y}))$ is false.
It is easy to see that $h(\alpha) \subseteq I$.
Also, since $\phi(h(\bar{x}),h'(\bar{y}))$ is false
for any valuation $h'$, for each $h'$
there must be an atom $F(x,y,z)\in\beta$, such that  
$h'\circ h(F(x,y,z))$
is false,
that is either $h'\circ h(F(x,y,z))= F(0,0,0)\notin I_{\phi}$
or $h'$ assigns for some existentially quantified 
variables non boolean values. 
Consequently
the trigger $(\xi,h)$ is active on $I_{\phi}$.

For the other direction, 
suppose that there
exists a trigger $(\xi,h)$ which is 
active on $I_{\phi}$,
i.e.,  $h(\alpha) \subseteq I_{\phi}$
and $h'(\beta) \not \subseteq I$,  
for any extension $h'$ of $h$.
This means that 
for any such extension $h'$, either
$h'$ is not $\{0,1\}$-valuation,
or that the atom $F(0,0,0)$ is in $h'(\beta)$.
Thus the formula
$\exists \bar{x}\; \neg \big(\exists \bar{y} \, \phi(\bar{x},\bar{y})\big)$
is true.
$_{\blacksquare}$

\medskip
Note that the trigger existence relates to the oblivious and semi-oblivious 
chase variations,
whereas the active trigger existence relates to the standard and the core chase.
This means that the oblivious and semi-oblivious chase variations 
have the same complexity. 
This is not the case for the standard and the core chase,
because the core chase step applies all active triggers in parallel and 
also involves the core computation for the resulted instance. 
In \cite{DBLP:conf/pods/FaginKP03} it
is shown that computing the core involves a
{\sf DP}-complete decision problem.

\bigskip
\section{Chase termination questions}

Being able to decide whether a chase should be terminated
at a given step in the sequence does not mean that
we can decide whether the case ever will terminate.
The latter problem is undecidable in general,
as we will see in the second subsection.
However, the several chase variations have
different termination behavior,
so next we introduce some notions 
that will help to distinguish them.

\subsection{Termination classes}\label{Sec:terminationClasses}

Let $\star\in\{{\sf std},{\sf obl},{\sf sobl},{\sf core}\}$,
corresponding to the chase variations introduced in 
Section \ref{prelim}, 
and let $\Sigma$ be a set of tgd's.
If there exists a terminating $\star$-chase sequence 
with $\Sigma$ on $I$, 
we say that the
$\star$-chase terminates for {\em some} branch
on instance~$I$,
and denote this as $\Sigma\in \actie{I}$.
Here $\actie{I}$ is thus to be understood
as the class
of all sets of tgd's for which the $\star$-chase terminates
on some branch on instance $I$.
Likewise, $\actia{I}$ denotes
the class of all sets of tgd's for which the $\star$-chase 
with $\Sigma$ on $I$ terminates on {\em all} branches.
From the definition of the chase variations
it is easy to observe that any trigger
applicable by the standard chase step on an instance $I$
is also applicable by the semi-oblivious and oblivious 
chase steps on the same instance. 
Similarly all the triggers applicable by the
semi-oblivious chase step on an instance $I$
is also applicable by the oblivious chase step on
instance $I$. Note that the converse is not always true.
Thus
$ \octie{I} \; \subseteq \;\soctie{I}\; \subseteq \;\sctie{I}.$
It is also easy to verify that
$ \actie{I}  = \actia{I} $
for $\star\in\{{\sf obl},{\sf sobl}\}$,
that $ \sctia{I} \; \subseteq \; \sctie{I}$, 
and that
$\soctia{I} \; \subseteq \; \sctia{I} $.
The following propositions shows that these 
results can be strengthened by including strict inclusions:

\begin{proposition} 
For any instance $I$ we have:
\begin{eqnarray*}
& & \octia{I}
= 
\octie{I}
\; \subset \;
\soctia{I}
= 
\soctie{I}
\; \subset \;
\sctia{I}
\; \subset \; 
\sctie{I}.
\end{eqnarray*}
\end{proposition}

{\em Proof:} (Sketch)
For strict inclusion
$\octie{I} \; \subset \; \soctia{I}$
consider 
instance $I=\{ R(a) \}$
and
$\Sigma$ 
containing a single (tautological) dependency 
$ R(x) \rightarrow \exists y\; R(y)$.
It is easy to see that $\Sigma \in \soctia{I}$
but $\Sigma \notin \octie{I}$.
For the second strict inclusion
$\soctie{I} \; \subset \; \sctia{I}$,
consider instance $I=\{ S(a,a) \}$
and $\Sigma=\{ S(x,y) \rightarrow \exists z\; S(y,z) \}$.
Because $I \models \Sigma$
it follows that  $\Sigma \in \sctia{I}$.
On the other hand, 
the semi-oblivious 
chase will not terminate with $\Sigma$ on $I$, 
and thus $\Sigma \notin \soctie{I}$.
For the final strict inclusion
$\sctia{I}
\; \subset \; 
\sctie{I}$, 
let $I=\{ S(a,b), R(a) \}$
and 
$\Sigma=\{ S(x,y)\rightarrow \exists z\; S(y,z); \;\; R(x) \rightarrow S(x,x) \}$. 
It is easy to see that any standard chase sequence that
starts by firing the trigger based on the first tgd will not terminate,
whereas if we first fire the trigger based on the second tgd, 
the standard chase will terminate after one step.
$_{\blacksquare}$

\bigskip
The next question is whether a $\star$-chase
terminates on {\em all} instances for all or 
for some branches.
The corresponding classes of sets of tgd's are
denoted $\actaa$ and
$\actae$, respectively.
Obviously
$ \sctaa \; \subset \; \sctae $.
Similarly to the instance dependent 
termination classes, 
$ \octaa \; \subset \; \soctaa $
and
$ \actaa \; = \; \actae $,
for $\star\in\{{\sf obl},{\sf sobl}\}$.


Now can relate the oblivious, semi-oblivious and standard 
chase termination classes as follows:

\begin{theorem}\label{PROP:semiObliviousChaseTerm}

\begin{eqnarray*}
\octaa
\; = \;
\octae 
\; \subset \; 
\soctaa=\soctae 
\; \subset \; 
\sctaa 
\; \subset \; 
\sctae.
\end{eqnarray*}

\end{theorem}

The proof of this theorem is included in the Appendix.

\bigskip

Note that for any 
$\star\in\{{\sf std},{\sf obl},{\sf sobl},{\sf core}\}$,
and for any non-empty instance $I$, we have that 
$\actaa \subset \actia{I}$
and 
$\actae \subset \actie{I}$.

\bigskip
The termination of the oblivious chase can be related to the 
termination of the standard chase by using the
 {\em enrichment}
transformation, introduced in \cite{DBLP:conf/amw/GrahneO11}.
The enrichment
takes a tgd 
$
\xi = 
\alpha(\bar{x},\mathbf{y})\rightarrow 
\exists \bar{z}\;
\beta(\bar{x},\bar{z})
$ over schema $\mathbf{R}$
and converts it into tuple generating dependency 
$
\depenrich{\xi} = 
\alpha(\bar{x},\bar{y})\rightarrow 
\exists \bar{z}\;
\beta(\bar{x},\bar{z}),{H}(\bar{x},\bar{y}),
$
where ${H}$ is a new relational symbol
that does not appear in~$\mathbf{R}$.
For a set $\Sigma$ of tgd's 
the transformed set is 
$\enrich{\Sigma} = \{\depenrich{\xi} : \xi\in\Sigma\}.$
Using the enrichment notion the following
was shown.

\begin{theorem}\label{THEO:oblivious-and-hat}
{\em \hspace*{-1ex}\cite{DBLP:conf/amw/GrahneO11}}
$\Sigma\in \octaa$ if and only if 
$\enrich{\Sigma} \in \sctaa$. 
\end{theorem}

To relate the termination of the semi-oblivious 
chase to the standard chase termination,
we use a transformation similar to the enrichment.
This transformation is called 
{\em semi-enrichment} and 
takes a tuple generating dependency
$
\xi = 
\alpha(\bar{x},\bar{y})\rightarrow 
\exists \bar{z}\;
\beta(\bar{x},\bar{z})
$ over a schema $\mathbf{R}$
and converts it into the tgd
$
\tilde{\xi} = 
\alpha(\bar{x},\bar{y})\rightarrow 
\exists \bar{z}\;
\beta(\bar{x},\bar{z}),{H}(\bar{x}),
$
where ${H}$ is a new relational symbol
which does not appear in~$\mathbf{R}$.
For a set $\Sigma$ of tgd's defined on schema $\mathbf{R}$,
the transformed set is 
$\widetilde{\Sigma} = \{\tilde{\xi} : \xi\in\Sigma\}.$
Using the semi-enrichment notion, 
the standard and the semi-oblivious chase 
can be related as follows.

\begin{theorem}\label{THEO:semi-oblivious-and-hat}
$\Sigma\in \soctaa$ if and only if
$\widetilde{\Sigma} \in \sctaa$. 
\end{theorem}

{\em Proof}: 
Similar to the proof of Theorem
\ref{THEO:oblivious-and-hat}.
$_{\blacksquare}$

\bigskip
We now turn our attention to the core chase.
Note the core chase is deterministic
since all active triggers are fired
in parallel, 
before taking the
core of the result. Thus we have: 

\begin{proposition}
$\cctia{I} = \cctie{I}$
and 
$\cctaa = \cctae$.
\end{proposition}

\medskip
It is well known that all here considered chase
variations
compute a (finite) {\em universal model}
of $I$ and $\Sigma$
when they terminate 
\cite{DBLP:conf/icdt/FaginKMP03,
DBLP:conf/pods/DeutschNR08,
DBLP:conf/kr/CaliGK08,
DBLP:conf/pods/Marnette09}.
In \cite{DBLP:conf/pods/DeutschNR08}, 
Deutsch et al.\ 
showed that if $I\cup\Sigma$ has a universal model,
the core chase will terminate in an instance that
is the core of all universal models. We thus have

\begin{proposition}\label{THEO:coreChaseTerminVSStandTermin}
$ $
\begin{enumerate}
\item
$\sctie{I} \, \subset \, \cctia{I}$,
for any instance~$I$.
\item
$\sctae \; \subset \; \cctaa$.
\end{enumerate}
\end{proposition}

{\em Proof:} (Sketch)
To see that the inclusion in part 2 of the proposition
is strict,
let $\Sigma = \{ {R}(x) \rightarrow \exists z\; {R}(z){S}(x)\}$, 
and 
${I_0}=\{ {R}(a) \}$. 
In this setting 
there will be exactly one 
active trigger at each step, 
and the algorithm will converge 
only at the infinite instance 
$$
\bigcup_{i>1} \{ R(z_i), S(z_{i-1}) \} \cup \{ R(a),S(a), R(z_1) \}.
$$
From this, it follows that $\Sigma \notin \sctaa$ and $\Sigma \notin \sctae$.
Note that for any positive integer $i$, the core of the instance ${I}_i$
is $\{ {R}(a),{S}(a) \}$. 
Thus the core chase will 
terminate at instance 
$I_1 = \{ {R}(a),{S}(a) \}$.
$_{\blacksquare}$

\bigskip

The following Corollary  highlights the 
relationship between the termination classes.

\begin{corollary}\label{COR:concl}
$\octaa \subset \soctaa \subset \sctaa \subset \sctae \subset \cctaa\mbox{.}$
\end{corollary}

\subsection{Undecidability of termination}

It has been known for some time that
``chase termination is undecidable.''
Specifically,
the following results have been obtained
in the literature so far.

\begin{theorem}\label{hitherto}
$ $
\begin{enumerate}
\item
$\mathsf{CT}^{\sf std}_{I, \forall}$ and 
$\mathsf{CT}^{\sf std}_{I, \exists}$
are {\sf RE}-complete
{\em \cite{DBLP:conf/pods/DeutschNR08}}.
\item
$\mathsf{CT}^{\sf core}_{I, \forall}
=
\mathsf{CT}^{\sf core}_{I, \exists}$,
and both sets are {\sf RE}-complete
{\em \cite{DBLP:conf/pods/DeutschNR08}}. 
\item
$\mathsf{CT}^{\sf sobl}_{I, \forall}=\mathsf{CT}^{\sf sobl}_{I, \exists}$,
and both sets are {\sf RE}-complete
{\em \cite{DBLP:conf/pods/Marnette09}}.
\item
Let $\Sigma$ be a set of guarded tgd's 
{\em \cite{DBLP:conf/kr/CaliGK08}}.
Then the question $\Sigma\in\mathsf{CT}^{\sf core}_{I, \forall}$
is decidable
{\em \cite{DBLP:conf/icdt/Hernich12}}.
\end{enumerate}
\end{theorem}

Our aim is to provide a systematic
overview on the complexity of all
termination classes.
We shall first show that 
$\mathsf{CT}^{\sf core}_{\forall\forall}$
is {\sf coRE}-complete.
To achieve this,
we provide a uniform reduction 
for both the
$\mathsf{CT}^{\sf core}_{I,\forall}$ and
$\mathsf{CT}^{\sf core}_{\forall\forall}$
problems, thus reproving
Theorem \ref{hitherto}, part 2
as a side effect.
We also note that the proofs in
\cite{DBLP:conf/pods/DeutschNR08} 
rely on machine reductions, and that Marnette 
observes in \cite{DBLP:phd/uk/Marnette10}
that a proof using a higher level
problem, such as Post's correspondence problem,
is still lacking. 
We fill this gap
using word-rewriting systems 
as redact.
A word-rewriting system, 
also known as semi-Thue systems,
is a set of rules of the form
$\ell\rightarrow r$, where $\ell$ and $r$
are words over a finite alphabet $\Delta$. 
Let $u$
and $v$ be words in $\Delta^*$.
The $u$ can be {\em derived} from $v$
if $u=x\ell y$ and $v=xry$,
for some $x,y\in\Delta^*$ and rule
$\ell\rightarrow r$.
A word rewriting system is terminating
for a word $w$ if the derivation closure of
$w$ is finite. The system is uniformly terminating
if it is terminating for all words $w\in\Delta^*$.
We will  prove that for every word rewriting
system and word there is a set of dependencies and an instance
such that
the rewriting system is terminating for that word if and
only if the core chase with the corresponding dependencies
on the corresponding instance terminates.
We also prove that if the core chase with the
corresponding dependencies is infinite on
all instances, then the rewriting system
is uniformly terminating.
It has long been known that
testing if a word-rewriting system terminates for 
a given input word is {\sf RE}-complete
\cite{david58}, and that 
testing if a  word-rewriting system
is uniformly terminating
is {\sf coRE}-complete \cite{huetLankford}.

\bigskip 
The next theorem states the main result of this section.

\begin{theorem}\label{THE:mainUndecidability}
The membership problem for $\cctaa$ is {\sf coRE}-complete.
\end{theorem}

The proof involves a development using a series of lemmas,
and can be found in its entirety in the Appendix. 
Our reduction from word rewriting systems also
yields

\begin{corollary}
The membership problem for
$\mathsf{CT}^{\sf core}_{I, \forall}$
is {\sf RE}-complete
(cf.\ {\em \cite{DBLP:conf/pods/DeutschNR08}}).
\end{corollary}

The same reduction from 
word-rewriting systems gives the following undecidability result too:

\begin{theorem}\label{THE:mainUndecidabilitySTD}
The membership problem for $\sctae$ is {\sf coRE}-complete.
\end{theorem}

Unfortunately the reduction used for the previous
results can't be used to show the undecidability 
of the $\sctaa$, $\soctaa$ or $\octaa$ classes.
To overcome this
we have allow a single {\em denial constraint},
that is a tgd of the form
$\alpha\!\rightarrow\!\bot$,
which is satisfied by an instance $I$
only if there is no homomorphism $h$,
such that $h(\alpha)\not\subseteq I$.
It is an open problem if the following
result (or part of it) can be obtained without
such constraints.

\begin{theorem}\label{THEO:sctaaCoRE}
In case the set of dependencies may contain 
at least one denial constraint, 
the membership problems for 
$\sctaa$, $\soctaa$ and $\octaa$ are {\sf coRE}-complete.
\end{theorem}

\bigskip
As a final observation we need to mention
that the class ``TOC'' of mappings defined
in \cite{DBLP:conf/pods/Marnette09},
for which termination of the semi-oblivious class
is proved to be {\sf RE}-complete,
is not the same with the class $\soctaa$.
Also there is no direct reduction from 
{\sf TOC} to $\soctaa$ 
membership problem, as former is defined 
only for sets of tgd's describing
{\em data exchange mappings},
and the question is whether the chase terminates for 
all instances that are source instances 
for the data exchange setting.

\section{Guaranteed termination}\label{sufficient}


To overcome the undecidability of chase termination,
a flurry restricted classes of tgd's have been 
proposed in the literature. These classes have
been put forth as subsets of
$\mathsf{CT}^{\sf std}_{\forall\forall}$,
although at the time only $\mathsf{CT}^{\sf core}_{I,\forall}$
was known to be undecidable. 
In this section we review these 
restricted classes
with the purpose of determining 
their overall structure and
termination properties.

Before reviewing these classes of sets of tgd's
let us define two properties attached to such classes
based on the enrichment and semi-enrichment rewritings
defined in subsection \ref{Sec:terminationClasses}.

A  class of sets of tgd's $\mathscr{C}$
is said to be {\em closed under enrichment}
if $\Sigma \in \mathscr{C}$
implies that $\enrich{\Sigma} \in \mathscr{C}$.
Using this notation together with Theorem \ref{THEO:oblivious-and-hat}
gives us a sufficient condition for a class of dependencies
to belong to $\octaa$:

\begin{proposition}\label{PROP:suffobl}
Let $\mathscr{C} \subseteq \sctaa$ such that $\mathscr{C}$
is closed under enrichment. 
Then
$\mathscr{C} \subseteq \octaa$.
\end{proposition}

Using this proposition we will reveal classes of 
dependencies that ensure termination for the 
oblivious chase.
Similarly we define the notion of {\em semi-enrichment closure} 
for classes of dependency sets.
The semi-enrichment closure property
gives a sufficient condition for the semi-oblivious chase
termination.

\begin{proposition}\label{PROP:suffsobl}
Let $\mathscr{C} \subseteq \sctaa$ such that $\mathscr{C}$
is closed under semi-enrichment. 
Then
$\mathscr{C} \subseteq \soctaa$.
\end{proposition}

As we will see next,
most of the known classes that ensure the 
standard chase termination are closed under semi-enrichment,
and thus those classes actually 
guarantee the semi-oblivious chase termination
as well. 
As we saw in Section \ref{SEC:complexityChaseStep}, 
the semi-oblivious chase has a lower complexity that
the standard chase.

\bigskip
\subsection*{Acyclicity based classes}

\bigskip
As full tgd's do not generate any new nulls
during the chase, any sequence with a set of
full tgd's will terminate
since there only is a finite number of tuples
that can be formed out of the elements of the
domain of the initial instance. The cause of
non-termination lies in the existentially
quantified variables in the head of the
dependencies. Most restricted classes thus
rely on restricting the tgd's in a way that
prevents these existential variables
to participate in any recursion.

The class of {\em weakly acyclic} sets
of tgd's 
\cite{DBLP:conf/icdt/FaginKMP03}
was one of the
first restricted classes to be proposed. 
Consider 
\begin{eqnarray*}
\Sigma_1 & = & \{R(x,y)\rightarrow\exists{z}S(z,x)\}.
\end{eqnarray*}

Let $(R,1)$ denote the first position in $R$,
and $(S,2)$ the second position in $S$, and so on.
In a chase step based on this dependency the values
from position $(R,1)$ get copied into
the position  $(S,2)$, whereas the value in position
$(R,1)$
``cause'' the generation of a new null value in $(S,1)$. 
This structure can been seen in the {\em dependency graph}
of $\Sigma_1$ that has a ``copy'' edge from vertex 
$(R,1)$ to vertex $(S,2)$,
and a ``generate'' edge from vertex $(R,1)$ to vertex $(S,1)$.
Note that the graph does not consider any
edges from $(R,2)$ because variable $y$ does not contribute
to the generated values.
The chase will terminate since there is no recursion
going through the $(S,2)$ position. By contrast, 
the dependency graph of 
\begin{eqnarray*}
\Sigma_2 & = & \{R(x,y)\rightarrow\exists{z}\;R(y,z)\}
\end{eqnarray*}
has a generating edge from
$(R,2)$ to $(R,2)$. It is the generating self-loop 
at $(R,2)$ which causes the chase on for example the instance
$\{R(a,b)\}$ to converge only at
the infinite instance 
$\{R(a,b),R(b,z_1)\}\cup\{R(z_i,z_{i+1}) :  i=1,2,\ldots\}$.
The class of weakly acyclic tgd's ({\sf WA}) is defined to be
those sets of tgd's whose dependency graph doesn't have
any cycles involving a generating edge
\cite{DBLP:conf/icdt/FaginKMP03}.
It is easy to observe that 
the class ({\sf WA}) is closed under 
semi-enrichment
but it is not closed under enrichment.
This is because in the case of semi-enrichment 
the new relational symbol $H$ considered 
for each dependency contains only variables
that appears both in the body and the head of the 
dependency,
and the new $H$ atoms appear only in the heads of the 
semi-enriched dependency.
This means 
that the dependency graph for 
a semi-enriched set of {\sf WA}  tgd's
will only add edges oriented into positions
associated with the new relational symbol.
The set $\Sigma=\{ R(x,y) \rightarrow \exists z\; R(x,z) \}$
shows that this is not the case for enrichment as
$\Sigma \in {\sf WA}$ but $\enrich{\Sigma} \notin {\sf WA}$.

The slightly smaller class of sets of tgd's
with {\em stratified witness} ({\sf SW}) 
\cite{DBLP:conf/icdt/DeutschT03} 
was introduced
around the same time as {\sf WA}.
An intermediate class, the {\em richly acyclic} 
tgd's ({\sf RA}) was introduced in 
\cite{DBLP:conf/pods/HernichS07} 
in a different context 
and it was later shown in
\cite{DBLP:conf/amw/GrahneO11}
that ${\sf RA}\in\mathsf{CT}^{\sf obl}_{\forall\forall}$.
It can be easily verified that both 
classes {\sf SW} and {\sf RA} are closed under 
enrichment.
The {\em safe dependencies} ({\sf SD} 
\cite{DBLP:journals/pvldb/MeierSL09}),
and the {\em super-weakly acyclic} ({\sf sWA} 
\cite{DBLP:conf/pods/Marnette09})
ones are both generalizations of the {\sf WA} class,
and both are close under semi-enrichment.

All of these classes have been proven to have {\sf PTIME}
membership tests, and have the following properties.

\begin{theorem}
{\em \cite{DBLP:conf/icdt/DeutschT03,DBLP:conf/icdt/FaginKMP03,DBLP:journals/pvldb/MeierSL09,DBLP:conf/pods/Marnette09,DBLP:conf/amw/GrahneO11}}
\begin{enumerate}
\item
${\sf SW} \; \subset \; {\sf RA} \; \subset \; {\sf WA}
\; \subset \; {\sf SD} \; \subset \; {\sf sWA}.$
\item
${\sf WA} \; \subset \; \mathsf{CT}^{\sf std}_{\forall\forall}$,
${\sf RA} \; \subset \; \mathsf{CT}^{\sf obl}_{\forall\forall}$, 
and
${\sf sWA} \; \subset \; \mathsf{CT}^{\sf sobl}_{\forall\forall}$. 
\end{enumerate}
\end{theorem}

\bigskip 
In order to 
complete the picture suggested by the previous
theorem we need a few more results.
Consider 
\begin{eqnarray*}
\Sigma_3 & =  & \{ {R}(x,y) \rightarrow \exists z\; {R}(x,z) \}.
\end{eqnarray*} 
Clearly $\Sigma_3 \in{\sf WA}$.
Let $I_0=\{ R(a,b) \}$,
and consider a semi-oblivious chase sequence
$I_0,I_1,I_2,\ldots$.
It is easy to see that for any $I_i, i>0$, 
there 
exists a (non-active) trigger $(\xi,\{ x/a, y/z_{i} \})$,
meaning that the oblivious chase will not terminate.
Thus we have $\Sigma_3 \notin \octaa$.
On the other hand, for the set
\begin{eqnarray*} 
\Sigma_4 & = & \{ {S}(y),{R}(x,y)\rightarrow \exists z\; {R}(y,z)\},
\end{eqnarray*}
we have 
$\Sigma_4 \in \octaa$.
Furthermore,
$\Sigma_4 \notin{\sf WA}$,
since the dependency graph of $\Sigma_4$ will
have a generating self-loop on vertex $(R,2)$. 
This gives us

\begin{proposition}\label{PROP:WAvsSemiOblivious}
The classes ${\sf WA}$ and $\octaa$ are incomparable
wrt inclusion.
\end{proposition}

It was shown in \cite{DBLP:journals/pvldb/MeierSL09} 
that  ${\sf WA} \subset {\sf SD}$ and also that 
${\sf SD} \subset \sctaa$. We can now extend this result by showing
that, similarly to the {\sf WA} class,  the following holds:

\begin{proposition}\label{PROP:SDSemiOblivious}
The classes ${\sf SD}$ and $\octaa$ are incomparable
wrt inclusion.
\end{proposition}

{\em Proof}: (Sketch)
The proof consists of showing that
$\Sigma_3\in{\sf SD}\setminus\octaa$,
and that $\Sigma_5 \in \octaa \setminus {\sf SD}$,
where 
\begin{eqnarray*}
\Sigma_5 & = & \{ R(x,x) \rightarrow \exists y\; R(x,y) \}.
\end{eqnarray*} 
Details are omitted. $_{\blacksquare}$

\medskip
From the semi-enrichment closure of the 
{\sf WA} and {\sf SD} 
classes and Proposition \ref{PROP:suffsobl}
we get the following result.

\begin{proposition}\label{PROP:soblTermClasses}
${\sf WA} \in \soctaa$
and ${\sf SD} \in \soctaa$.
Furthermore,
for any instance $I$ and any $\Sigma\in {\sf SD} $ 
the semi-oblivious chase 
with $\Sigma$ on $I$ terminates in time polynomial in the size
of $I$.
\end{proposition}

Note that the previous result follows directly also from 
a similar result for the class {\sf sWA}
 \cite{DBLP:conf/pods/Marnette09}.
Still, as shown by the following proposition,
the super-weakly acyclic 
class does not include the class of dependencies that ensures 
termination for the oblivious chase variation, nor does
the inclusion hold in the other direction.

\begin{proposition}\label{PROP:SWAnotOCTAA}
${\sf sWA}$ and $\octaa$ are incomparable wrt inclusion.
\end{proposition}

{\em Proof}: (Sketch)
We exhibit the super-weakly acyclic set
$\Sigma_3=\{ {R}(x,y) \rightarrow \exists z\; {R}(x,z) \}$. 
It is clear $\Sigma_3 \notin \octaa$.
For the converse, let
\begin{eqnarray*}
\Sigma_6 & = & \{{S}(x),{R}(x,y) \rightarrow \exists z\;{R}(y,z) \}.
\end{eqnarray*} 
Then $\Sigma_6 \notin {\sf sWA}$,
on the other hand it can be observed that the oblivious chase
with $\Sigma_6$
terminates on all instances. This is because
tuples with new nulls cannot cause the dependency to fire,
as these new nulls will never be present in relation $S$.$_{\blacksquare}$



\bigskip 
\bigskip 
\subsection*{Stratification based classes}

\bigskip
Consider $\Sigma_7 = \{\xi_{1},\xi_{2}\}$, where
\begin{eqnarray*}
\xi_{1} & = & R(x,x)  \rightarrow  \exists{z}\;S(x,z), \mbox{ and}  \\ 
\xi_{2} & = & R(x,y),S(x,z)  \rightarrow  R(z,x).
\end{eqnarray*}

In the dependency graph of $\Sigma_7$ we will
have the cycle $(R,1) \rightsquigarrow (S,2) \rightsquigarrow (R,1)$,
and since $(S,2)$ is an existential position,
the set $\Sigma_7$ is not weakly acyclic.
However, it is easy to see that
$\Sigma_7\in\mathsf{CT}^{\sf std}_{\forall\forall}$.
It is also easily seen that if $S$ is empty and $R$ non-empty,
then $\xi_1$ will ``cause'' $\xi_2$ to fire 
for every tuple in $R$. Let us denote this
relationship by $\xi_1\prec \, \xi_2$.
On the other hand,
there in no chase sequence in which
a new null in $(S,2)$ can be propagated
back to a tuple in $R$ and to fire a trigger based on $\xi_1$,
and thus create an infinite loop. 
We denote this with
$\xi_2\not\prec\xi_1$.
In comparison,
when chasing with 
\begin{eqnarray*}
\Sigma_8 & = & \{R(x,y)\rightarrow\exists{z}\;R(z,x)\}, 
\end{eqnarray*}
the new
null $z_i$ in $(z_{i},z_{i-1})$ will propagate into
tuple $(z_{i+1},z_{i})$, in an infinite regress.
If we denote the tgd in $\Sigma_8$ with $\xi$,
we conclude that $\xi\prec \, \xi$.
A formal definition of the 
$\prec$ relation is given in the
next section.

The preceeding observations led Deutsch et al.\ 
\cite{DBLP:conf/pods/DeutschNR08} to 
define the class of
{\em stratified} dependencies
by considering the {\em chase graph} 
of a set $\Sigma$,
where the individual tgd's in $\Sigma$ are the
vertexes and there is an edge from $\xi_1$ to $\xi_2$
when $\xi_1\prec \, \xi_2$. A set $\Sigma$ is then
said to be stratified if the vertex-set of
every cycle in the chase graph forms a weakly
acyclic set. The class of all sets of stratified tgd's
is denoted {\sf Str}.
In the previous example,
$\Sigma_7\in{\sf Str}$, and
$\Sigma_8\notin{\sf Str}$.

In \cite{DBLP:journals/pvldb/MeierSL09}
Meier et al.\ 
observed that 
${\sf Str} \not \subseteq \sctaa$ and that
actually only 
${\sf Str}\subset\mathsf{CT}^{\sf std}_{\forall\exists}$,
and came up with a corrected definition of
$\prec$, which yielded the
{\em corrected stratified} class
{\sf CStr} of tgd's,
for which they showed

\begin{theorem}
{\em \cite{DBLP:journals/pvldb/MeierSL09}}
$$ 
{\sf CStr}\subset \sctaa,\;
{\sf Str} \subset \sctae,\; and \;
{\sf CStr}\subset~{\sf Str}.
$$
\end{theorem}

\bigskip
From the observation that the 
 {\sf CStr} class is closed under semi-enrichment
and from Proposition \ref{PROP:suffsobl}
we have:

\begin{proposition}\label{PROP:cstrinSemiOblivious}
$ $
\begin{enumerate}
\item
${\sf CStr}\subset \soctaa$.
\item
${\sf CStr}$ and $\octaa$ are incomparable wrt
inclusion.
\end{enumerate}
\end{proposition}

{\em Proof}: (Sketch)
For the second part we have $\Sigma_3 \in {\sf CStr}$
and $\Sigma_3 \notin \octaa$. For the converse
consider the dependency set $\Sigma_6$ from the proof of
Proposition \ref{PROP:SWAnotOCTAA}.$_{\blacksquare}$

\medskip
Meier et al.\ \cite{Meier2009} further observed that the
basic stratification definition
also catches some false negatives.
For this they considered
dependency set $\Sigma_9 = \{\xi_3,\xi_4\}$, where

\vspace{-0.6cm}
\begin{eqnarray*}\label{EQ:oblivNonStandardTermination}
\xi_3 & = & S(x),E(x,y) \rightarrow  {E}(y,x), \mbox{ and}\\
\xi_4 & = & S(x),E(x,y) \rightarrow \exists z\; E(y,z),E(z,x). 
\end{eqnarray*}

Here $\xi_3$ and $\xi_4$ belong to the same
stratum according to the definition of {\sf CStr}.
Since new nulls in both $(E,1)$ and $(E,2)$
can be caused by $(E,1)$ and $(E,2)$,
there will be generating self-loops on
these vertexes in the dependency graph.
Hence $\Sigma_9\notin{\sf CStr}$. On the other
hand, it is easy to see that the number of
new nulls that can be generated in the chase 
is bounded by the number of tuples in
relation $S$ in the initial instance.
Consequently $\Sigma_9\in\mathsf{CT}^{\sf std}_{\forall\forall}$.

In order to avoid such false negatives,
Meier et al.\ \cite{Meier2009} gave an alternative
definition of the $\prec$ relation
and of the chase graph.
Both of theses definitions are however
technically rather involved, and will
not be repeated here. 
The new {\em inductively restricted} class,
abbreviated  {\sf IR},
restricts each connected component 
in the modified chase graph
to be in {\sf SD}. 
In example above, 
$\Sigma_9 \in {\sf IR}$.

Meier et al.\ \cite{Meier2009} also observed that {\sf IR}
only catches {\em binary}
relationships $\xi_1\prec \, \xi_2$.
This could be 
generalized to a ternary relation $\prec(\xi_1,\xi_2,\xi_3)$,
meaning that there exists a chase sequence
such that firing $\xi_1$ will cause $\xi_2$ to fire, 
and this in turn causes $\xi_3$ to fire. 
This will eliminate those cases
where $\xi_1$, $\xi_2$ and $\xi_3$ form a connected
component in the (modified) chase graph,
and yet there is no chase sequence that will fire
$\xi_1$, $\xi_2$ and $\xi_3$ in this order.
Thus the tree dependencies should not be in the same stratum.

Similarly to the ternary extension, 
the $\prec$ relation can be generalized to be $k$-ary.
The resulting termination classes are denoted 
${\sf T}[k]$. Thus ${\sf T}[2] = {\sf IR}$,
and in general ${\sf T}[k]\subset{\sf T}[k+1]$ 
\cite{Meier2009}. The main property is

\begin{theorem}
{\em \cite{DBLP:journals/pvldb/MeierSL09}}
$$
 {\sf CStr}
\subset {IR} = {\sf T}[2]
\subset {\sf T}[3]
\subset \cdots
\subset {\sf T}[k]
\subset \cdots
\subset\mathsf{CT}^{\sf std}_{\forall\forall}.
$$
\end{theorem}

To complete the picture, we have the following
proposition based on the semi-oblivious closure for the
{\sf T}[k] hierarchy and Proposition \ref{PROP:suffsobl}.

\begin{proposition}\label{PROP:cstrinSemiOblivious}
$ $
\begin{enumerate}

\item
${\sf T}[k] \subset\soctaa$.
\item
${\sf T}[k]$ and  $\octaa$ are incomparable
wrt inclusion. 
\end{enumerate}
\end{proposition}

Before concluding this section need to mention
that all the classes discussed here are closed under semi-enrichment,
thus they ensure the termination for the less expensive 
semi-oblivious chase in a polynomial number of steps, 
in the size of the input instance.

The Hasse diagram in Figure 2 from the Appendix 
summarizes
the stratification based classes and
their termination properties.

\bigskip
\section{Complexity of stratification}\label{complexity}

\bigskip
As we noted in Section \ref{sufficient},
all the acyclicity based classes
have the property that testing whether
a given set $\Sigma$ belongs to it
can be done in {\sf PTIME}. The situation
changes when we move to the stratified classes.
The authors of 
\cite{DBLP:conf/pods/DeutschNR08}
claimed that testing if $\xi_1 \prec \,\xi_2$ is in {\sf NP} for a 
given $\xi_1$ and $\xi_2$,
thus resulting in {\sf Str} having a {\sf coNP} membership problem.
We shall see in Theorem \ref{coNPhard} below
that this cannot be the case,
unless ${\sf NP}={\sf coNP}$.
We shall use the $\prec$ order as it is defined
for the {\sf CStr} class. The results also hold
for the {\sf Str} class. 
First we need a formal
definition.

\begin{definition}\label{meier}{\em \cite{DBLP:journals/pvldb/MeierSL09}}
Let $\xi_1$ and $\xi_2$ be tgd's.
Then $\xi_1$ {\em precedes} $\xi_2$,
denoted
$\xi_1 \prec \, \xi_2$, 
if there exists an instance $I$ and homomorphisms
$h_1$ and $h_2$ from the universal variables in
$\xi_1$ and $\xi_2$, respectively,
such that:
\begin{list}{(\roman{foo})~}{\usecounter{foo}}
\item 
$I\models h_2(\xi_2)$, and
\item
$I\xrightarrow{(\xi_1,h_1)} J$ using an oblivious chase step,and
\item 
$J \not \models 
h_2(\xi_2)$.
\end{list}
\end{definition}
Note that the pair $(\xi_1,h_1)$ in the previous 
definition denotes a trigger, not necessarily an active trigger,
because the chase step considered is the oblivious one.
Intuitively, the instance $I$ in the definition
is a witness to the ``causal'' relationship
between $\xi_1$ and $\xi_2$
(via $h_2$),
as $h_2(\xi_2)$ won't fire at $I$,
but will fire once $\xi_1$
has been applied.
The notion of stratum of $\Sigma$ is as before, 
i.e.\ we build a chase graph consisting of
a vertex for each tgd in $\Sigma$,
and an edge from $\xi_1$ to $\xi_2$
if $\xi_1\prec \, \xi_2$.
Then  
$\xi_1$ and $\xi_2$ are
in the same {\em stratum} when they both
belong to the same cycle
in the chase graph of $\Sigma$. 
A set $\Sigma$ of tgd's is said to be 
{\em C-stratified} ({\sf CStr}) if all
its strata are weakly acyclic
\cite{DBLP:journals/pvldb/MeierSL09}.

\begin{theorem}\label{coNPhard}
$ $ 
\begin{enumerate}
\item
Given two tgd's $\xi_1$ and $\xi_2$, the problem of
testing if $\xi_1 \prec \, \xi_2$ is {\sf coNP}-hard.
\item
Given a set of dependencies $\Sigma$,
the problem of  testing if $\Sigma \in {\sf CStr}$ 
is {\sf NP}-hard.
\end{enumerate}
\end{theorem}

{\bf Proof}:
For part 1 of the theorem
we will use a reduction from the graph 3-colorability 
problem that is known to be {\sf NP}-complete.
It is also well known that a graph $G$ is 3-colorable iff there is 
a homomorphism from $G$ to $K_3$,
where $K_3$ is the complete graph with 3 vertices.
We provide a reduction $G\mapsto\{\xi_1,\xi_2\}$,
such that $G$ is not 3-colorable if and only if
$\xi_1\prec \, \xi_2$.

We identify a graph 
$G=(V,E)$, where $|V|=n$ and $|E|=m$ with the sequence
$$
G(x_1,\ldots,x_n) = E(x_{i_{1}},y_{i_{1}}),\ldots,E(x_{i_{m}},y_{i_{m}}),
$$ 
and treat the elements
in $V$ as variables.  
Similarly, we identify the graph
$K_3$ with the sequence $K_3(z_1,z_2,z_3) =$ 
$$E(z_1,z_2),E(z_2,z_1),E(z_1,z_3),E(z_3,z_1),
E(z_2,z_3),E(z_3,z_2)$$ 
where $z_1,z_2$, and $z_3$ are variables.
With these notations, 
given a graph $G=(V,E)$, 
we construct tgd's $\xi_1$
and $\xi_2$ as follows:

\vspace{-0.6cm}
\begin{alignat*}{2}
\xi_1 &= \;\;\;\; R(z)  & \rightarrow\;& \exists z_1,z_2,z_3\; K_3(z_1,z_2,z_3), \mbox{ and} \\
\xi_2 &=  E(x,y)\;& \rightarrow\;& \exists x_1,\ldots,x_n\;G(x_1,\ldots,x_n).
\end{alignat*}

Clearly the reduction is polynomial in the size of $G$.
We will now show that $\xi_1 \prec \,\xi_2$ iff
$G$ is not 3-colorable.

First, suppose that 
$\xi_1 \prec \,\xi_2$.
Then there exists 
an instance $I$ and homomorphisms $h_1$ and $h_2$,
such that $I \models h_2(\xi_2)$.
Consider $J$, where
$I \xrightarrow{(\xi_1,h_1)} J$. 
Thus $R^I$ had to contain at least 
one tuple, and $E^I$ had to be empty,
because otherwise the monotonicity 
property of the chase we would imply that
that $J \models h_2(\xi_2)$.

On the other hand, 
we have 
$I \xrightarrow{(\xi_1,h_1)} J$,
where instance $J = I \cup \{K_3(h_1'(z_1),h_1'(z_2),h_1'(z_3))\}$,
and $h_1'$ is a distinct extension of $h_1$.
Since $E^I=\emptyset$, and we assumed that
$J \not \models h_2(\xi_2)$,
it follows that 
there is no homomorphism from 
$G$ into $J$,
i.e.\ 
there is no homomorphism from 
$G(h'_2(x_1),\ldots,h'_2(x_n))$ to 
$K_3(h_1'(z_1),h_1'(z_2),h_1'(z_3))$,
where $h'_2$ is a distinct extension of $h_2$.
Therefore the graph $G$
is not 3-colorable.

For the other direction, let us suppose that graph $G$ is not 3-colorable.
This means that there is no homomorphism from $G$ into $K_3$. 
With these assumption let us consider $I=\{ R(a) \}$
homomorphism $h_1=\{ z/a \}$ and homomorphism 
$h_2=\{ x/h'_1(z_1), y/h'_1(z_2) \}$.
It is easy to verify that $I$, $h_1$ and $h_2$
satisfy the three
conditions for $\xi_1 \prec \, \xi_2$.

\medskip
For part 2 of the theorem,
consider the set $\Sigma=\{ \xi_1, \xi_2 \}$ 
defined as follows:

\begin{eqnarray*}
&\xi_1 \; =& \; R(z_1,v)\;  \rightarrow \exists z_2,z_3,w\; K_3(z_1,z_2,z_3), \\
& &\ \ \ \ \ \ \ \ \ \ \ \ \ \ \ \ R(z_2,w), R(z_3,w), S(w), \mbox{ and} \\
& \xi_2 \; =& \; E(x,y)\; \rightarrow  \exists x_1,\ldots,x_n,v\; G(x_1,\ldots,x_n),R(x,v).
\end{eqnarray*}

It is straightforward to verify
that $\Sigma \notin {\sf WA}$ and 
that $\xi_2 \prec \, \xi_1$. 
Similarly to the proof of part 1,
it can be shown that $\xi_1 \prec \, \xi_2$ iff the graph $G$ is not 
3-colorable. From this follows that $\Sigma \in {\sf CStr}$
iff there is no cycle in the chase graph,
iff $\xi_1 \not \prec \,\xi_2$ iff 
$G$ is 3-colorable. $_{\blacksquare}$

\bigskip
Note that the 
reduction in the previous proof
can be used to show that  the 
problem $\Sigma \in {\sf Str}$ is {\sf NP}-hard.
Similar result can be also obtained
for the {\sf IR} class and also for the 
{\em local stratification} based classes introduced by Greco et al. in 
\cite{DBLP:journals/pvldb/GrecoST11}.
The obvious upper bound for the problem $\xi_1 \prec \,\xi_2$ 
is given by:

\bigskip
\begin{proposition}\label{easy}
Given two dependencies $\xi_1$ and $\xi_2$, 
the problem of determining whether
$\xi_1 \prec \, \xi_2$ is 
in $\Sigma_2^p$.
\end{proposition}

{\bf Proof}:
From \cite{DBLP:conf/pods/DeutschNR08} 
we know that if $\xi_1 \prec \, \xi_2$
there is an instance $I$ satisfying 
Definition \ref{meier}, such that 
size of $I$ is bounded by a polynomial in the size of 
$\{\xi_1,\xi_2\}$. 
Thus, we can guess instance~$I$, homomorphisms $h_1$ and 
$h_2$ in {\sf NP} time. 
Next, with a {\sf NP} oracle 
we can check if 
$I \models h_2(\xi_2)$ and 
$J \not \models h_2(\xi_2)$, where
$I \xrightarrow{(\xi_1,h_1)} J$. $_{\blacksquare}$

\bigskip
We shall see that the upper bound of the proposition
actually can be lowered to $\Delta^p_2$.
For this we need the following characterization theorem. 

\begin{theorem}\label{THEO:charac}
Let $\xi_1=\alpha_1 \rightarrow \beta_1$ and 
$\xi_2=\alpha_2 \rightarrow \beta_2$ be tgd's. Then,
$\xi_1 \prec \, \xi_2$ if and only if
there is an atom $t$, and homomorphisms
$h_1$ and $h_2$, such that the following hold.
\begin{list}{(\alph{qcounter})~}{\usecounter{qcounter}}
\item 
$t \in \beta_1$,
\item 
$h_1(t) \in h_2(\alpha_2)$, 
\item 
$h_1(t) \notin h_1(\alpha_1)$, and
\item There is no idempotent homomorphism from
$h_2(\beta_2)$ to
$h_2(\alpha_2) \cup h_1(\alpha_1) \cup h_1(\beta_1)$.
\end{list}
\end{theorem}

{\bf Proof}:
We first prove the ``only if`` direction. 
For this, suppose that
$\xi_1 \prec \, \xi_2$, that is, 
there exists an instance
$I$ and homomorphisms 
$g_1$ and $g_2$,
such that conditions
$(i) - (iii)$ of Definition \ref{meier}
are fulfilled.

From conditions $ii$ and $iii$ 
we have that
$g_1(\alpha_1)  \subseteq I$ and 
$g_1'(\beta_1) \not \subseteq I$,
for any distinct extension $g_1'$ of $g_1$. 

Now, 
consider $h_1=g_1$ and $h_2=g_2$.
Let $t$ be an atom from $\beta_1$ such that 
$h'_1(t) \in h'_1(\beta_1) \cap h_2(\alpha_2)$  and 
$h'_1(t) \notin h_1(\alpha_1)$, for an extension $h'_1$ of $h_1$. 
Such an atom $t$ must exists, since
otherwise it will be that  
$h'_1(\beta_1) \cap h_2(\alpha_2) \subseteq h_1(\alpha_1)$,
which is not possible
because of conditions $(i)$ and $(iii)$ (note that $h_1=g_1$).
It is now easy to see that $t$, $h'_1$ and $h_2$ satisfy conditions
$(a), (b)$, and $(c)$ of the theorem.
It remains to show that condition $(d)$ also is satisfied.
By construction
we have $J= I \cup h'_1(\beta_1)$.
It now follows
that $I \cup h'_1(\beta_1) \not \models h_2(\xi_2)$.
Because $h_1(\alpha_1) \subseteq I$,
condition $(d)$  is indeed satisfied.

\medskip 
For the ``if'' direction of the theorem,
suppose that there exists an atom $t$
and homomorphisms $h_1$ and $h_2$,
such that 
conditions $(a),(b),(c)$ and $(d)$ holds.
Let $g_1=h_1$, $g_2=h_2$
and let $I=(h_1(\alpha_1) \cup h_2(\alpha_2)) \setminus h'_1(t)$,
for a distinct extension $h'_1$ of $h_1$.
Because $h'_1(t) \notin I$ and $h'_1(t) \in h_2(\alpha_2)$,
it follows that $h_2(\alpha_2) \not \subseteq I$.
Thus we have $I \models h_2(\xi_2)$, proving 
point $(i)$ of Definition
\ref{meier}. 
On the other hand, 
because point $(c)$ of the theorem is assumed, 
it follows that
$h_1(\alpha_1) \subseteq I$, 
from which we get
$I \xrightarrow{(\xi_1,h_1)} J$, 
where $J = I \cup h'_1(\beta_1)$,
proving  points $(i)$ and $(ii)$ from Definition \ref{meier}.  
Since
$I \cup h'_1(\beta_1)=h_1(\alpha_1) \cup h_2(\alpha_2) \cup h'_1(\beta_1)$,
and point $(d)$ holds, we get
$J \not \models h_2(\xi_2)$,
thus showing that condition $(iii)$ of Definition \ref{meier}
is also satisfied. $_{\blacksquare}$

\bigskip
It is easy to note that by adding the extra 
condition
{\em $(e)$  there is no idempotent homomorphism from $\beta_1$ to $\alpha_1$}
in the previous theorem we obtain a characterization of the 
stratification order associated with the {\sf Str} class.

With this characterization result we can now 
tighten the
$\Sigma_2^p$ upper bound of Proposition \ref{easy}
as follows:

\begin{theorem}
Given two dependencies $\xi_1$ and $\xi_2$,
the problem of determining whether
$\xi_1 \prec \, \xi_2$ is 
in $\Delta_2^p$.
\end{theorem}

{\bf Proof}:
For this proof we will use the 
characterization Theorem \ref{THEO:charac},
and the observation  that 
$\Delta_2^{p} = {\sf P}^{\sf NP}={\sf P}^{\sf coNP}$.
Consider the following {\sf PTIME} algorithm that 
enumerates all possible $h_1, h_2$ and $t$:

\begin{codebox}
\li \For $t \in \beta_1$ 
\li \Do \For all $(h_1,h_2)=\mbox{mgu}(t,\alpha_2)$
\li     \Do \If $h_1(t)\notin h_1(\alpha_1)$ \Return $t, h_1, h_2$ \End \End
\end{codebox}

\noindent 
In the algorithm, 
$\mbox{mgu}(t,\alpha_2)$ denotes all 
pairs $(h_1,h_2)$
such that there exists an atom $t' \in \alpha_2$,
with $h_1(t)=h_2(t')$,
and there is no $(g_1,g_2)$ 
and $f$ different from the identity mappings, 
such that
$h_1=g_1 \circ f$
and 
$h_2=g_2 \circ f$.

\noindent
Using the values returned by previous algorithm and
with a {\sf coNP} oracle 
we can test if point $(d)$
holds.
Thus, the problem is in $\Delta_2^p$. $_{\blacksquare}$

Armed with these results we can now state the 
upper-bound for the complexity 
of the {\sf CStr} 
membership problem.

\begin{theorem}\label{THEO:cstrUpperBound}
Let $\Sigma$ be a set of tgd's, then
the problem of testing if $\Sigma\in {\sf CStr}$ 
is in $\Pi_2^p$.
\end{theorem}

\noindent
{\bf Proof}: 
(Sketch)
To prove that $\Sigma$ is not in {\sf CStr}
guess a set of tuples $(\xi_1,t^1,h^1_1,h^1_2)$,
$\ldots$,$(\xi_k,t^k,h^k_1,h^k_2)$,
where $\xi_1, \ldots, \xi_k $ are tgd's in $\Sigma$,
$t_1,\ldots,t_k$ are atoms, and $h^i_1$,$h^i_2$
are homomorphisms, for $i \in \{ 1,\ldots,k \}$.
Then, using an ${\sf NP}$ oracle 
check that 
$\xi_i < \xi_{i+1}$, for $i \in \{1,\ldots, k-1 \}$, 
and $\xi_k < \xi_1$, using
the characterization Theorem \ref{THEO:charac}
with $t^i$, $h^i_1$, $h^i_2$ and
$t^k$, $h^k_1$, $h^k_2$ respectively.
And then check in  {\sf PTIME}
if the set of dependencies $\{ \xi_1, \ldots, \xi_k \}$
is not weakly acyclic.
Thus, the complexity is $\Pi_2^P$. $_{\blacksquare}$

\bigskip
We note that using the obvious upper-bound $\Sigma_2^p$ for 
testing if $\xi_1 \prec \xi_2$, the membership problem for 
the class {\sf CStr} would be in $\Pi_3^p$.
As mentioned the same results apply also for the class {\sf Str}.
Even if the complexity bounds for testing if $\xi_1 \prec \; \xi_2$
are not tight, it can be noted that a {\sf coNP} upper bound 
would not lower the $\Pi_2^P$ upper bound of the membership problem
for {\sf CStr}.

\bigskip
\section{Conclusions}

\bigskip
We have undertaken a systematization of the
somewhat heterogeneous area of the chase.
Our analysis produced a taxonomy of the 
various chase versions and their termination
properties,
showing that the main sufficient classes 
that guarantee termination for the standard chase
also ensures termination for the complexity-wise less 
expensive semi-oblivious chase.
Even if the standard chase procedure 
in general captures more sets of dependencies 
that ensure the chase termination than the 
semi-oblivious chase we argue that for most
practical constraints 
the semi-oblivious chase is a better choice.
We have also proved that the membership
problem for the classes $\cctaa$ and $\sctae$
is {\sf coRE}-complete
and in case we allow also at least one 
denial constraint the 
same holds for  $\sctaa$, $\soctaa$ and $\octaa$.
Still it remains an if the membership problem 
for  $\sctaa$ remains  {\sf coRE}-complete
without denial constraints.
The same also holds
for the classes $\soctaa$ and $\octaa$.
Finally we have analyzed the
complexity of the membership problem
for the class of stratified sets of dependencies.
Our bounds for this class are not tight, and 
it remains an open problem to pinpoint
the complexity exactly.

\bigskip
\bibliographystyle{abbrv}

\bibliography{GrahneOnet}

\newpage
\appendix
\bigskip

Note that the theorems presented in the paper we kept the same numbering 
and the new theorems/lemmas are numbered continuously.

\section*{4. Chase termination questions}

\setcounter{theorem}{1}
\begin{theorem}\label{PROP:semiObliviousChaseTerm}

\begin{eqnarray*}
\octaa
\; = \;
\octae 
\; \subset \; 
\soctaa=\soctae 
\; \subset \; 
\sctaa 
\; \subset \; 
\sctae.
\end{eqnarray*}

\end{theorem}

{\bf Proof}: (Sketch)
We will only show the strict inclusion parts of the theorem.
For the first inclusion, let
$\Sigma=\{ R(x,y) \rightarrow \exists z\; R(x,z) \}$.
It is easy to see that $\Sigma\in \soctaa$ and 
$\Sigma\notin \octae$.
The second strict inclusion 
$\soctaa \; \subset \; \sctaa$ 
is more intricate, 
as most sets of  dependencies in $\sctaa$ part of $\soctaa$.
To distinguish the two classes,
let $\Sigma = \{\xi_1,\xi_2\}$, where

\vspace{-0.1cm}
\begin{eqnarray*}\label{EQ:oblivNonStandardTermination}
\xi_1 & = & {R}(x) \rightarrow \exists z\; {S}(z),{T}(z,x),\mbox{ and}\\
\xi_2 & = & {S}(x) \rightarrow \exists z'\; {R}(z'),{T}(x,z').
\end{eqnarray*}

\vspace{0.1cm}
Let now $I$ be an arbitrary instance,
and suppose that $I = \{R(a_1),\ldots,R(a_n),S(b_1),\ldots,S(b_m)\}$.
There is no loss of generality,
since if the standard chase with $\Sigma$ on $I$ terminates, 
then it will also terminate even if the initial instance
contains atoms over the relational symbol $T$.
It is easy to see that all standard chase sequences 
with $\Sigma$ on $I$
will terminate in the instance 
$A\cup B$, where

\vspace{-0.1cm}
\begin{eqnarray*}
A & = & \bigcup_{i\in\{1,\ldots,n\}} \{R(a_i),S(z_i),T(z_i,a_i)\},\mbox{ and}   \\
B & = & \bigcup_{i\in\{1,\ldots,m\}} \{R(z'_i),S(b_i),T(b_i,z'_i)\}.
\end{eqnarray*}

\vspace{0.1cm}
On the other hand, all semi-oblivious chase
sequences will converge only at the infinite instance
$$
\bigcup_{k\in\mathbb{N}} \big(C_k \cup D_k\big) \cup A \cup B,$$
where

\vspace{-0.1cm}
{\small{
\begin{equation*}
\begin{split}
C_k =&\bigcup_{i=1}^{m} \{ S(z_{kn+(k-1)m+i}), 
T(z_{kn+(k-1)m+i},z'_{(k-1)m+(k-1)n+i}) \} \cup                 \\
   &   \bigcup_{j=1}^{n} \{S(z_{kn+km+j}), T(z_{kn+km+j},z'_{km+(k-1)n+j}) \},                             \\
D_k=& \bigcup_{i=1}^{n} \{ R(z'_{km+(k-1)n+i}),T(z_{(k-1)n+(k-1)m+i},z'_{km+(k-1)n+i}) \} \cup                 \\
   &\bigcup_{j=1}^{m} \{R(z'_{km+kn+j}),T(z_{kn+(k-1)m+j},z'_{km+kn+j}) \}.
\end{split}
\end{equation*}
}}

For the last inclusion $\sctaa \subset \sctae$,  consider the set
$\Sigma=\{ R(x,y)\rightarrow R(y,y);\;\; R(x,y)\rightarrow \exists z\; R(y,z) \}$. 
Clearly $\Sigma \in \sctae$, 
because for any instance 
all chase sequences that start by firing the first tgd will terminate.
On the other hand  $\Sigma \notin \sctaa$, 
as the standard chase with $\Sigma$
does not terminate on
$I=\{ R(a,b) \}$
whenever
the second tgd is applied first.
$_{\blacksquare}$

\section*{4.2 Undecidability of termination}

\bigskip
Hithereto, the following results have been obtained.

\setcounter{theorem}{4}
\begin{theorem}\label{A:hitherto}
$ $
\begin{enumerate}
\item
$\mathsf{CT}^{\sf std}_{I, \forall}$ and 
$\mathsf{CT}^{\sf std}_{I, \exists}$
are {\sf RE}-complete
{\em \cite{DBLP:conf/pods/DeutschNR08}}.
\item
$\mathsf{CT}^{\sf core}_{I, \forall}
=
\mathsf{CT}^{\sf core}_{I, \exists}$,
and both sets are {\sf RE}-complete
{\em \cite{DBLP:conf/pods/DeutschNR08}}. 
\item
$\mathsf{CT}^{\sf sobl}_{I, \forall}=\mathsf{CT}^{\sf sobl}_{I, \exists}$,
and both sets are {\sf RE}-complete
{\em \cite{DBLP:conf/pods/Marnette09}}.
\item
Let $\Sigma$ be a set of guarded tgd's 
{\em \cite{DBLP:conf/kr/CaliGK08}}.
Then the question $\Sigma\in\mathsf{CT}^{\sf core}_{I, \forall}$
is decidable
{\em \cite{DBLP:conf/icdt/Hernich12}}.
\end{enumerate}
\end{theorem}

\medskip
Before describing the reduction and the proofs for
 theorems \ref{THE:mainUndecidability},  \ref{THE:mainUndecidabilitySTD} 
and \ref{THEO:sctaaCoRE}
let us first give a brief description for the word-rewriting systems.

\medskip
\noindent
{\bf Word rewriting systems}.
Let $\Delta$ be a finite set of symbols,
denoted $a,b,\ldots$, possibly
subscripted,
and $\Delta^*$ the set of all
finite words over~$\Delta$.
Let $\Theta$ be a finite subset of
$\Delta^*\!\times\Delta^*$. 
Treating each pair in $\Theta$ as a {\em rule},
the relation $\Theta$ gives rise to
a {\em rewriting relation}
$\rightarrow_{\Theta} \;\; \subseteq \;\; \Delta^*\!\times\Delta^*$
defined~as
$
\{(u,v)
\; : \; u = x\ell y, v = xry,\;
(\ell,r)\in\Theta\}. 
$

We use the notation $u\rightarrow_{\Theta} v$ 
instead of $\rightarrow_{\Theta}\!(u,v)$.
If $\Theta$ is understood from the context we will simply
write $u\rightarrow v$.
If we want to emphasize
which rule $\rho\in\Theta$ was used we write
$u\rightarrow_{\rho} v$.
When $u\rightarrow_{\rho} v$ we say that
$v$ is obtained from $u$ by  a {\em rewriting step}
(based on $\rho$).  
By $u \rightarrow^n v$ we mean that
$v$ can be obtained from $u$ in 
at most $n$ rewriting steps. 
A {\em rewriting system} is then a pair
$(\Delta^*,\Theta)$. If $\Delta$
is understood
from the context we shall
denote a rewriting system simply with~$\Theta$.

A sequence $w_0,w_1,w_2,\ldots$ of words from
$\Delta^*$ is said to be a 
$\Theta$-{\em derivation sequence}
(or simply a derivation sequence),
if $w_i\rightarrow w_{i+1}$ for 
all $i=0,1,2,\ldots$.
A derivation sequence might be finite or infinite. 
The {\em termination problem for $\Theta$ and a word} 
$w_0\in\Delta^*$,
is to determine whether
all derivation sequences
$w_0,w_1,w_2,\ldots$ originating from $w_0$
are finite. 
The {\em uniform termination problem for}
$\Theta$ is to determine
whether for {\em all} words $w_0\in\Delta^*$,
it holds that all derivation sequences
$w_0,w_1,w_2\ldots$ originating from $w_0$
are finite. It has long been known that
the termination problem is {\sf RE}-complete
\cite{david58},
and that the uniform termination problem
is {\sf coRE}-complete \cite{huetLankford}.

\medskip

We now describe our reduction 
$\Theta\mapsto\Sigma_{\Theta}$.
We assume without loss of generality that $\Delta=\{0,1\}$.
The tgd set $\Sigma_{\Theta}$ is defined for schema
$\mathbf{R}_{\Theta} = \{E,E^*,L,R,D\}$ and consists of 
$\{\xi_{\rho} : \rho\in\Theta\}
\cup \{ \xi_{L_{0}}, \xi_{L_{1}}, \xi_{R_{0}}, \xi_{R_{1}} \} \cup TC \cup AD
\cup S$, where $\xi_{\rho}$ is:
{\small{
\begin{eqnarray*}
& & \hspace{-0.6cm} E(x_0,a_1,x_1),\ldots,E(x_{n-1},a_n,x_n) \; \rightarrow \exists \; y_0,\ldots,y_m \; L(x_0,y_0),\\
& & \hspace{1.5cm} E(y_0,b_1,y_1),\ldots,E(y_{m-1},b_m,y_m), R(x_n,y_m).
\end{eqnarray*}
}}
when $\rho = (a_1\ldots a_n, b_1\ldots b_m)$.
We will also have the following ``grid creation'' rules, 
using left $L$ and right $R$ predicates.

{\small{
\begin{eqnarray*}
\xi_{L_{0}} & = & E(x_0,0,x_1),L(x_1,y_1)  \; \rightarrow \;\;\; \exists \; y_0\; L(x_0,y_0), E(y_0,0,y_1) \\
\xi_{L_{1}} & = & E(x_0,1,x_1),L(x_1,y_1)  \; \rightarrow \;\;\; \exists \; y_0\; L(x_0,y_0), E(x_0,1,y_1) \\
\xi_{R_{0}} & = & R(x_0,z_0), E(x_0,0,x_1) \; \rightarrow \;\;\; \exists \; z_1\; E(z_0,0,z_1),R(x_1,z_1)  \\
\xi_{R_{1}} & = & R(x_0,z_0), E(x_0,1,x_1) \; \rightarrow \;\;\; \exists \; z_1\; E(z_0,1,z_1),R(x_1,z_1) 
\end{eqnarray*}
}}

In the sequel we will assume,
unless otherwise stated, that
all instances are over schema $\mathbf{R}_{\Theta}$.
For an instance~$I$ 
the following rule set $AD$ (``active domain'')
computes  $\dom(I)\cup\Delta$ 
in relation $D$.

{\small{
\begin{eqnarray*}
            &\rightarrow& \; D(0), D(1)      \\
E(x,z,y) \; &\rightarrow& \; D(x),D(z),D(y)  \\
L(x,y)   \; &\rightarrow& \; D(x),D(y)       \\
R(x,y)   \; &\rightarrow& \; D(x),D(y)       \\
E^*(x,y) \; &\rightarrow& \; D(x),D(y) 
\end{eqnarray*}
}}

Given an instance $I$ we denote by
$G_I$ the graph with edge set:

{\small{
$$
\{ (x,y)\;:\; E(x,z,y) \in I \mbox{ or } E^*(x,y) \in I 
\mbox{ or }  L(x,y)\in I \mbox{ or }  R(x,y)\in I \}.
$$
}}

The following set $TC$ 
computes in $E^*$ the transitive closure
of $G_I$: 

{\small{
\begin{eqnarray*}
E(x,z,y) \;           &\rightarrow& \;  E^*(x,y) \\
L(x,y) \;             &\rightarrow& \;  E^*(x,y) \\
R(x,y) \;             &\rightarrow& \;  E^*(x,y) \\
E^*(x,y),E^*(y,z) \;  &\rightarrow& \;   E^*(x,z) 
\end{eqnarray*}
}}

If an instance has a cycle in $E^*$,
that is there exists a cycle in $G_I$,
the dependencies in
the following ``saturation'' set $S$
will be fired: 

{\small{
\begin{eqnarray*}
E^*(v,v),D(x),D(z),D(y) \; &\rightarrow& \; E(x,z,y) \\
E^*(v,v),D(x),D(y) \;      &\rightarrow& \; L(x,y),R(x,y),E^*(x,y) 
\end{eqnarray*}
}}

In the sequel, we shall
sometimes say that an instance $I$ is cyclic (acyclic)
if $G_I$ is cyclic (acyclic). We shall also speak of 
``the graph of $I$'', when we mean $G_I$.

We denote by $H_I$ the Herbrand base of instance $I$,
i.e.\ the instance where, for each relation symbol $R\in\mathbf{R}_{\Theta}$,
the interpretation
$R^{H_{I}}$ contains all tuples (of appropriate arity) that can be formed
from the constants in 
$(\dom(I)\cap{\sf Cons})\cup\Delta$.
The proof of the following lemma is straightforward:

\begin{lemma}\label{LEMMA:herbrandCore}
$\core(I) = H_I$, whenever 
$H_I$ is a subinstance of $I$. 
\end{lemma}

\begin{lemma}\label{LEMMA:cycleTermination}
Let $I$ be an arbitrary instance over schema $\mathbf{R}_{\Theta}$,
and let $I=I_0$,$I_1$,$I_2$,$\ldots$  be the core chase sequence 
with $\Sigma_{\Theta}$ on $I$. 
If there is an integer $i$ and a
constant or variable $x$, such that $E^*(x,x)\in I_i$ 
(i.e.\ the graph $G_{I_{j}}$ is cyclic
for some $j\leq i$), then the core chase sequence is finite. 
\end{lemma}

\noindent 
{\em Proof:}  (Sketch)
First we note that $H_{I_{k}} = H_{I}$
for any instance $I_k$ in the core chase sequence,
since the chase does not add any new constants.
If the core chase does not terminate at
the instance $I_i$ mentioned in the claim,
if follows that the dependencies in the set $S$
will fire at $I_i$ and generate $H_{I}$ as a subinstance.
It then follows from Lemma \ref{LEMMA:herbrandCore}
that $I_{i+1} = H_{I}$.
It is easy to see that
$H_{I}\models\Sigma_{\Theta}$,
so the core chase will terminate at 
instance~$I_{i+1}$.
$_{\blacksquare}$

\bigskip
Intuitively the previous lemma guarantees that 
whenever we have a cycle in the initial instance 
the core chase process will terminate. Thus,
in the following we will not have to care about instances
that may contain cycles.

\bigskip
The following lemma ensures  that 
the core chase with $\Sigma_{\Theta}$
on an acyclic instance will not create
any cycles.

\begin{lemma}\label{LEMMA:noCycle}
Let $I$ be an arbitrary acyclic instance over schema 
 $\mathbf{R}_{\Theta}$,
and let  $I=I_0,I_1,I_2,\ldots$ be the core chase
sequence with $\Sigma_{\Theta}$ on $I$. 
Then $G_{I_{i}}$ is acyclic,
for all instances $I_i$
in the sequence.
\end{lemma}

{\em Proof:}  (Sketch)
Suppose to the contrary that $G_{I_{i}}$ is cyclic,
for some $I_i$ in the sequence.
Wlog we assume that $I_i$ is the first such instance
in the sequence.
Clearly $i\geq\!{1}$.
This means that by applying all 
active triggers on $I_{i-1}$
will add a cycle 
(note that the taking the core cannot add a cycle).
Let $(\xi_1,h_1),\ldots, (\xi_n,h_n)$
be the triggers that add tuples to $I_i$, 
causing  $G_{I_{i}}$ to be cyclic.
First, it is easy to see 
that 
$\{ \xi_{L_{0}},\xi_{L_{1}}, \xi_{R_{0}}, \xi_{R_{1}}  \} 
\cap 
\{ \xi_1, \ldots, \xi_n \} = 
\emptyset$. 
This is because these dependencies 
do not introduce any new edges in $G_{I_{i}}$
between vertices in $G_{I_{i-1}}$,
they only add a new vertex into $G_{I_{i}}$ which
will have two incoming edges from vertices
already in $G_{I_{i-1}}$.
A similar reasoning shows 
that none of the $\xi_{\rho}\in TC$ or $\xi\in AD$
can be part of the set $\{ \xi_1, \ldots, \xi_n \}$.
Finally, the dependencies in the set $S$ may introduce 
cycles and may thus be part of the set 
$\{ \xi_1, \ldots, \xi_n \}$.
But the dependencies in $S$ are fired only when
$E^*(x,x)\in I_{i-1}$,
which means that 
$G_{I_{i-1}}$ already contains a cycle,
namely the self-loop on $x$.
Contradicts our counter assumption that
$I_{i}$ is the first instance 
in the chase sequence that contains a cycle.
$_{\blacksquare}$

\bigskip 
We still need a few more notions.
A {\em path} $\pi$ of an instance~$I$ over $R_{\Theta}$
is a set 
$$\{E(x_0,a_1,x_1),E(x_1,a_2,x_2),\ldots,E(x_{n-1},a_n,x_n)\}$$ 
of atoms of~$I$, 
such that $\{a_1, a_2, \ldots, a_n\} \subseteq  \Delta$ 
(recall that $\Delta$ is the alphabet of 
the rewriting system $\Theta$).
The word spelled by the path $\pi$ is
$$\word(\pi) = a_1a_2\ldots a_n.$$
A~{\em max-path} $\pi$ in an instance 
$I$ is a path, such that no path $\pi'$ in $I$
is a strict superset of $\pi$.
We can now relate words and instances as follows:
let $I$ be an acyclic instance,  we define
$$\paths(I) = \{\pi : \pi \mbox{ is a max-path in } I\}.$$
Clearly $\paths(I)$ is finite, for any finite instance $I$.
Conversely, let $w=a_1a_2\ldots a_n \in \Delta^*$. We define
$$I_w=\{ E(x_0, a_1, x_1), E(x_1, a_2, x_2), \ldots, E(x_{n-1}, a_n, x_n) \},$$ 
where the $x_i$'s are pairwise distinct variables.
Clearly $\paths(I_w) = \{\pi\}$, where
$\word(\pi) = w$.

\begin{lemma}
Let $w \in \Delta^*$, 
and let $I_w=I_0, I_1, I_2, \ldots$  
be the core chase sequence with 
$\Sigma_{\Theta}$ on $I_w$. For each instance
$I_i$ in the sequence, denote by $I'_i$ the instance
obtained from $I_i$ by firing all active triggers,
that is $I_{i+1}=\core(I'_i)$.
Then $\paths(I'_i)=\paths(I_{i+1})$.
\end{lemma}

{\em Proof:}  (Sketch)
If there were a path $\pi$,
such that $\pi\in\paths(I'_i)\setminus \paths(\core(I'_i))$,
there would have to be atoms of the form
$L(x,x)$ and $R(x,x)$ in the instance $I'_i$,
which would contradict Lemma \ref{LEMMA:noCycle}
as $I_w$, by definition, does not contain any cycles.
$_{\blacksquare}$

\bigskip

In order to be able to relate rewrite sequences
and core chase sequences
we introduce rewrite trees
and 
path trees.
For a rewriting system $\Theta$ and word $w \in \Delta^*$,
we construct the {\em rewrite tree} 
$\mathcal{T}_w$ inductively.
Start with a root node labelled $w$.
Then, for each leaf node $\mathsf{n}$ in $\mathcal{T}_w$, 
for each possible derivation
$v\rightarrow u$,
where $v$ is the label of $\mathsf{n}$,
add a new node $\mathsf{m}$
as a child of $\mathsf{n}$,
and label $\mathsf{m}$ with $u$.


The next lemma follows directly from the
construction of $\mathcal{T}_{w}$.

\begin{lemma}\label{LEMMA:TWinfinite}
Let $w \in \Delta^*$. 
Then $\mathcal{T}_{w}$ 
has an infinite branch
if and only if there is
an infinite rewriting derivation
$w=w_0, w_1, w_2, \ldots$
generated by $\Theta$ from $w$. 
\end{lemma}


\bigskip

We next define 
the {\em path tree} $\mathcal{P}_w$
of the core chase sequence 
$I_w=I_0,I_1,I_2,\ldots$
generated by $\Sigma_{\Theta}$ from $I_w$.
The path tree is defined inductively 
on the levels of the tree.
First, let $\mathcal{P}_w$
consist of a single node,
labelled with the
single path in $\paths(I_w)$.
Inductively, for each leaf node $\mathsf{n}$ in $\mathcal{P}_w$,
where $\pi$ is the label of $\mathsf{n}$,
\begin{enumerate}
\item
for each $\xi_{\rho}=\alpha\!\rightarrow\!\beta\in\Sigma_{\Theta}$
if there is a homomorphism $h$, such that the trigger
$(\xi_{\rho},h)$ is active on $I_i$
and $h(\alpha)\subseteq\pi$,
then add a child $\mathsf{m}$ labelled with
the unique path in
$\paths(h'(\beta))$, 
where $h'$ is the distinct extension of $h$.
\item
for each pair $(\xi_L,\xi_R)$,
say $(\xi_{L_{0}},\xi_{R_{1}})$,
if there is a homomorphism $h$,
such that $(\xi_{L_{0}},h)$
is active on $I_i$
(and/or $(\xi_{R_{1}},h)$
is active on $I_i$), 
and $\pi$ contains
$E(h(x_0),0,h(x_1))$
(and/or
$E(h(x_0),1,h(x_1))\in\pi$),
let $h'$ be the distinct extension of $h$, and
add a child node 
$\mathsf{m}$, 
labelled with the unique path in the set 
$$\paths\big(E(h'(y_0),0,h'(y_1))\cup\pi\big)$$
(labelled with the unique path in
$$\paths\big(E(h'(y_0),0,h'(y_1))\cup\pi\cup E(h(x_0),1,h(z_1))\big)$$
or with the unique path in
$$\paths\big(\pi\cup E(h'(z_0),1,h'(z_1))\big),$$ respectively).
\end{enumerate}

Similarly to Lemma \ref{LEMMA:TWinfinite}, we have

\begin{lemma}\label{LEMMA:PWInfinite}
Let $w \in \Delta^*$. 
Then $\mathcal{P}_{w}$ is infinite
if and only if
the core chase sequence
$I_w=I_0, I_1, I_2, \ldots$
on $I_w$
with $\Sigma_{\Theta}$
is infinite.
\end{lemma}


\begin{figure*}\label{FIG:tree}
  \begin{center} 
\includegraphics[scale=0.33]{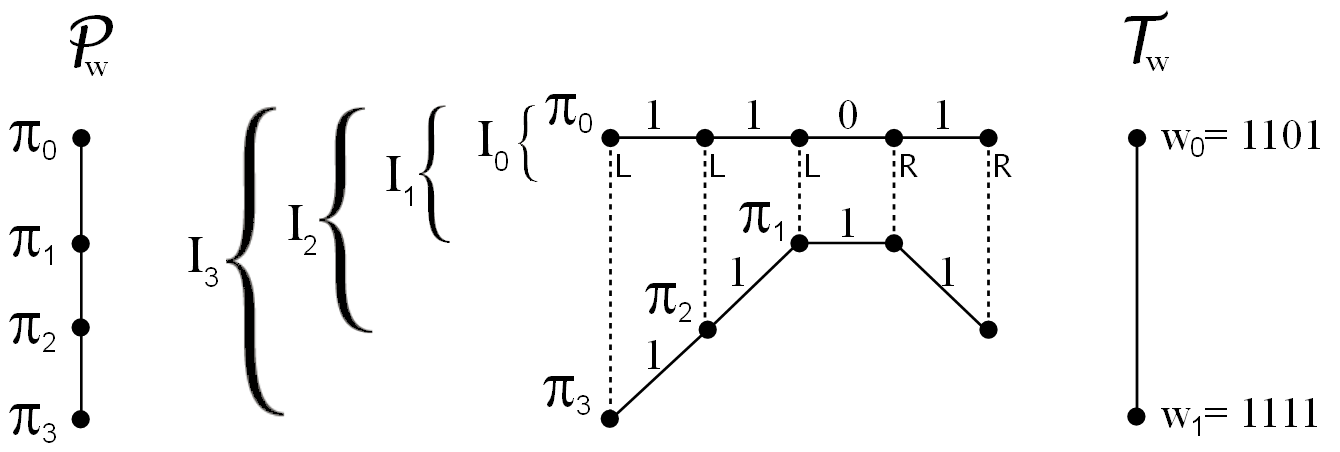}
 \caption{A simple reduction and the corresponding trees.}
  \end{center}
\end{figure*} 
\medskip
We can now state the following important theorem.

\setcounter{theorem}{15}
\begin{theorem}\label{ATHEO:reprove}
Let $w\in\Delta^*$.
Then the core chase sequence $I_w=I_0, I_1, I_2, \ldots$
with $\Sigma_{\Theta}$ on $I_w$ is infinite
if and only if 
there is an infinite derivation 
$w=w_0,w_1,w_2,\ldots $ 
generated by $\Theta$.
\end{theorem}

{\em Proof:}  (Sketch)
For the  if part, suppose that there is an infinite 
derivation $w=w_0,w_1,w_2,\ldots $.
From Lemma \ref{LEMMA:TWinfinite}
it follows that 
$\mathcal{T}_w$ has an infinite branch
and by construction
the branch is labelled
by the derivation sequence
$w=w_0, w_1, w_2, \ldots$ generated by $\Theta$.  
We claim that 
there is an infinite branch in
$\mathcal{P}_w$ 
labelled with
$\pi_0,\pi_1,\pi_2, \ldots$,
and a sequence of indices
$0=j_0<j_1<j_2<\cdots$,
such that $\word(\pi_{j_{i}}) = w_i$,
for all  $i=0,1,2,\ldots$.
Clearly $\word(\pi_0) = w_0$.
For the inductive hypothesis fix $n$, 
and suppose that
$\word(\pi_{j_{i}}) = w_i$,
for all $i=0,1,\ldots,n$.
Let $w_{n} = x\ell y$
and $w_{n+1} = xry$.
Also, let $k=max\{|x|,|y|\}$.
It follows from 
the inductive hypothesis and
the construction of
$\mathcal{P}_w$, 
that the branch labelled
$\pi_{j_{0}},\ldots,\pi_{j_{1}},\ldots,\pi_{j_{n}}$,
where $\word(\pi_{j_{i}}) = w_i$,
continues with nodes labelled
$\pi_{j_{n}+1}, \pi_{j_{n}+2}, \ldots, \pi_{j_{n}+k}$,
where $\word(\pi_{j_{n}+k}) = w_{n+1}$.
Since $\mathcal{P}_w$ thus has an infinite branch,
Lemma \ref{LEMMA:PWInfinite}
tells us that
the core chase sequence is 
infinite as well. 
Figure 1
shows the relationship between $\mathcal{P}_w$
and $\mathcal{T}_w$ where $\Theta=\{ (0,1) \}$
and the initial word $w_0$ is $1101$.

For the other direction,
suppose that the core chase sequence is 
infinite.
From Lemma \ref{LEMMA:PWInfinite}
it follows that 
 $\mathcal{P}_w$ has an infinite branch.
Let this branch be labelled
$\pi_0,\pi_1,\pi_2,\ldots$.
We claim that there is an infinite
derivation
$w=w_0, w_1, w_2, \ldots$ generated by $\Theta$,  
and a sequence
$0=j_0<j_1<j_2<\cdots$ of indices,
such that $\word(\pi_{j_{i}})=w_i$,
for all $i=0,1,2,\ldots$.
This can be seen by choosing
$j_{i+1}=j_{i}+k$, where
$\pi_{j_{i}+k}$ 
is the first label 
on the path in $\mathcal{P}_w$,
such that 
$\pi_{j_{i}}\not\subseteq\pi_{j_{i}+k}$.
$_{\blacksquare}$

\medskip
This theorem together with the
{\sf RE}-completeness result for
rewriting termination \cite{david58},
yields the undecidability result of
Deutsch et al. \cite{DBLP:conf/pods/DeutschNR08} 
for core chase termination on a given instance.

\begin{corollary}\label{ATHEO:DeutschTheo}
The set $\mathsf{CT}^{\sf core}_{I, \forall}$
is {\sf RE}-complete.
\end{corollary}

Next we shall relate the uniform termination
problem with the set $\cctaa$. 
This means that we need to consider arbitrary instances,
not just instances of the form $I_w$, for $w\in\Delta^*$.
First, we introduce a few more notations.

We denote by $\Sigma_{\rho}$
the set of all $\xi_{\rho}$ dependencies
in $\Sigma_{\Theta}$. 
Likewise,
$\Sigma_{LR}$ will denote the set
$\{ \xi_{L_{0}}, \xi_{L_{1}}, \xi_{R_{0}}, \xi_{R_{1}} \}$.

Let $I$ be an acyclic instance, and 
$
I'=\chase_{\Sigma_{LR}}^{\sf core}(I)$. 
Where 
$\chase_{\Sigma}^{\sf core}(I)$
represents the instance returned by the core chase procedure
on $I$ with $\Sigma$.
It is easy to see that $I'$ is finite
for any finite acyclic instance $I$.
Then let 
$I^* = \bigcup \{I_{\word(\pi)} : \pi\in\paths(I')\}$.
Intuitively, $I^*$ is obtained from $I$ by
by taking each max-path in $I'$,  
and making it into a unique line tree in $G_{I^{*}}$.

\begin{lemma}\label{iterLemma}
Let $I$ be an arbitrary acyclic instance.
Then, the core chase 
of $I$ with $\Sigma_{\Theta}$
terminates if and only if
the core chase of $\chase_{\Sigma_{LR}}(I)$ with 
$\Sigma_{\Theta}$ terminates.
\end{lemma}

{\em Proof:}
Let $I=I_0, I_1, \ldots$ be the core 
chase sequence of $I$ with $\Sigma_{\Theta}$
and $\chase_{\Sigma_{LR}}(I)=J_0, J_1, \ldots$
be the core chase sequence 
of $\chase_{\Sigma_{LR}}(I)$ with $\Sigma_{\Theta}$.
Let us first suppose that the core chase sequence on
$\chase_{\Sigma_{LR}}(I)$ does not terminate.
In this case it is easy to see that 
that there must be an integer $i$ such that 
$\chase_{\Sigma_{LR}}(I) \subseteq I_i$,
but from this follows that the core chase for $I$ with
$\Sigma_{\Theta}$ does not terminate either.
The other direction follows directly from the observation 
that for any $i$ we have $I_i \subseteq J_i$. $_{\blacksquare}$

\medskip
\begin{lemma}\label{istar}
Let $I$ be an arbitrary acyclic 
instance such that $I \models \Sigma_{LR}$.
Then the core chase of $I$ with $\Sigma_{\Theta}$
is infinite if and only if 
the core chase of $I^*$ with $\Sigma_{\Theta}$
is infinite.
\end{lemma}

{\em Proof:} (Sketch)
Let $I=I_0, I_1, I_2, \ldots$ 
be the core chase sequence of
$I$ with $\Sigma_{\Theta}$.
And let $I^*=J_0, J_1, \ldots$
be the core chase sequence of
$I^*$ with $\Sigma_{\Theta}$.
Suppose that the sequence 
$I_0, I_1, I_2, \ldots$ is infinite.

We will prove by induction
that for each $i$ there exists a $j$
such that 
for each path $\pi \in \paths(I_i)$ 
there exists a unique path 
$\pi' \in \paths(J_j)$
and $\word(\pi)$ is a factor of
$\word(\pi')$. This proving the if part of the lemma.

For the base case, let $i=0$ and consider $j=0$.
By the definition $I^*$
contains all the paths in $\paths(I)$.
For the inductive step let us suppose that for 
a fixed $i$ it holds that for
any integer $k\leq i$
there exists 
$j_k$ 
such that 
for each path $\pi \in \paths(I_k)$ 
there exists a unique path 
$\pi' \in \paths(J_{j_k})$
such that $\word(\pi)$ is a factor of
$\word(\pi')$.
For each path $\pi$ in 
$\pi \in \paths(I_{i+1}) \setminus \paths(I_{i})$
we will assign a unique 
path $\pi' \in \paths(J_j) \setminus \paths(J_{j_i})$ 
for some $j\geq j_i$ by considering the following 2 cases:

{\em Case 1}. $\pi$ was created by applying a $\xi_\rho$
dependency for some $\rho=(a_1\ldots a_n, b_1\ldots b_m) \in \Theta$.
In this case it needs to be that there exists
$\pi_0 \in \paths(I_{i})$
such that 
$a_1\ldots a_n$ is a factor of $\word(\pi_0)$.
From the induction hypothesis it follows that there exists 
a $j$ and a unique $\pi'_0 \in \paths(J_j)$
such that $\word(\pi_0)$ is a factor of $\word(\pi'_0)$.
By transitivity of word factor relation it follows that 
$a_1\ldots a_n$ is also a factor of $\word(\pi'_0)$.
But this means that  the same dependency $\xi_{\rho}$
can be applied, or was already applied, for
$I_j$, following that
$\paths(I_{j+1})$ contains the path $\pi'$
such that $\word(\pi)= b_1 \ldots b_m$ is a factor
of $\word(\pi')$.

{\em Case 2}. $\pi$ was created by extending path $\pi_1 \in \paths(I_i)$
using one or two dependencies from
$\Sigma_{LR}$. 
Because of the assumption $I \models \Sigma_{LR}$ 
it follows that there must exist a subpath $\pi_2$ of $\pi_1$
(note that such a subpath is unique)
such that $\pi_2$ was obtained from a $\xi_{\rho}$
dependency applied to path $\pi_3 \in \paths(I_k)$, where $k<i$
and $\rho=(a_1\ldots a_n, b_1\ldots b_m)$.
Thus, $\word(\pi_2)=b_1\ldots b_m$ and
$\word(\pi_3)=v_0 a_1 \ldots a_n v_1$, for some words $v_0$ and $v_1$.
Because the core computation
 does not shrink existing paths
it follows that there must be a path $\pi_4 \in \paths(I_i)$
such that $\pi_3$ is a subpath of $\pi_4$,
thus the spelling for $\pi_4$ will be
$\word(\pi_4)=u_0 v_0 a_1 \ldots a_n v_1 u_1$, for some words $u_0$ and $u_1$.
Based on the acyclic requirement for $I$,
 Lemma \ref{LEMMA:noCycle},
the definition for the $\Sigma_{LR}$ dependencies
and the observation that applying a $\xi_\rho$
dependencies will always create a new path
 it follows that 
$\word(\pi)$ is a factor of the word 
$u_0 v_0 b_1 \ldots b_m v_1 u_1$.
From the induction hypothesis we have that 
for $\word(\pi_4)=u_0 v_0 a_1 \ldots a_n v_1 u_1$
there exists a $j$ and a unique path  $\pi' \in \paths(J_j)$ 
such that $\word(\pi_4)$
is a factor of $\word(\pi')$.
This means that the core chase process 
can applied (or already applied) trigger
$(\xi_\rho,h)$ 
that mapped the body of $\xi_\rho$
to the path given by word $a_1 \ldots a_n$
in $\word(\pi_4)=u_0 v_0 a_1 \ldots a_n v_1 u_1$.
By applying this trigger instance $J_{j+1}$
will contain the path $\pi'_1$
that is a factor of $u_0 v_0 b_1 \ldots b_m v_1 u_1$.
After applying a maximum of $max(|u_0 v_0|,|v_1 u_1|)$
core chase steps (i.e.\ applying the copying dependencies)
it follows that instance $J_{j+1+max(|u_0 v_0|,|v_1 u_1|)}$
will contain a path $\pi'$
such that $u_0 v_0 b_1 \ldots b_m v_1 u_1$ is a factor 
of $\word(\pi')$. Note that the assignment of $\pi$ 
to $\pi'$ is unique due to the uniqueness of the trigger applied.
Also by the transitivity of the factor relation it follows that
$\word(\pi)$ is also a factor of $\word(\pi')$.

For the only if direction
let us suppose that the 
core chase of 
$I^*$ with $\Sigma_{\Theta}$
does not terminate.
Because all the paths in 
$\paths(I^*)$ does not share any node in common,
it follows that there must be a 
path $\pi$ in $\paths(I^*)$ such that 
the core chase of $I_{\word(\pi)}$
does not terminate.
Clearly from the definition of $I^*$
it follows that in the core chase sequence
$I=I_0,I_1, I_2, \ldots $
there exists and integer $i$
and there exists path $\pi' \in \paths(I_i)$
with $\word(\pi')=\word(\pi)$.
This means that for the instance corresponding with the path $\pi'$
the core chase will follow  
core chase steps isomorphic with the once
used when chasing instance 
$I_w$, where $w={\word(\pi)}$. 
Thus, the core chase for $\pi'$
will not terminate either. $_{\blacksquare}$

\medskip 
We can now state the following important result
\begin{theorem}\label{ATHEO:terminationForAll}
A reduction system $\Theta$ uniformly terminates if and only if
the core chase terminates on all instances for $\Sigma_{\Theta}$
(i.e. $\Sigma_{\Theta} \in \cctaa$).
\end{theorem}

{\em Proof:}  (Sketch)
First let us suppose that $\Sigma_{\Theta} \in \cctaa$.
Let $w \in \Delta^*$ be an arbitrary word.
Because $\Sigma_{\Theta} \in \cctaa$ it follows that 
the core chase will terminate also with instance $I_w$.
From this and Theorem \ref{ATHEO:reprove} it follows that 
the rewriting system $\Theta$ will terminate for $w$.

For the other direction
suppose that $\Sigma_{\Theta} \notin \cctaa$.
Then there exists an instance $I$,
such that the core chase sequence 
$I=I_0,I_1,I_2,\ldots$ 
with $\Sigma_{\Theta}$ 
is infinite.
From Lemma \ref{LEMMA:cycleTermination}
it must be that $I$ is acyclic.
From Lemma \ref{iterLemma} and 
Lemma \ref{istar}
it follows that the core chase of $I^*$
with $\Sigma_{\Theta}$
must be infinite as well.
But then there must be path $\pi\in I^{*}$
such that $\Theta$ admits an infinite derivation
starting from $\word(\pi)$.
But this means that $\Theta$ is not uniformly 
terminating.
$_{\blacksquare}$

\medskip

Using the previous result and 
the {\sf coRE}-completeness result
of Huet and Lankford \cite{huetLankford},
we now have the main theorem.

\setcounter{theorem}{5}
\begin{theorem}\label{ATHE:mainUndecidability}
The membership problem for $\cctaa$ is {\sf coRE}-complete.
\end{theorem}

\medskip
Up to here we proved the undecidability of the $\cctaa$ class.
Next we will show that this result can be extended with some minor changes
to other termination classes as well.

\begin{theorem}\label{ATHEO:sctaeUndecidable}
The membership problem for $\sctae$ is {\sf coRE}-complete.
\end{theorem}

{\em Proof:}  (Sketch)
It is easy to see that the same $\Sigma_{\Theta}$
reduction works for the $\sctae$ case as well 
by choosing the branch that first applies 
all the $AD\cup TC \cup S$ dependencies.
This is because in case the initial arbitrary instance contains a cycle
the full dependencies $AD\cup TC \cup S$ 
will saturate the instance and the standard chase will terminate.
If $I$ does not contain any cycles then, as we showed, during the 
chase process no cycles are added and the termination proof
is the same as for the core chase.
$_{\blacksquare}$

\medskip 
To show that the basic $\Sigma_{\Theta}$ reduction can't be used for 
the $\sctaa$ class. Consider the word-reduction system $\Theta=\{ (1,0) \}$
and instance $I=\{ E(a,0,a),L(a,b) \}$. It is easy to see that 
the branch that applies the $\xi_{L_{0}}$ dependency
first will not terminate as it will generate the following
infinite set of tuples:

\begin{minipage}[b]{0.50\linewidth}
\centering
\begin{tabular}{lll} 
\multicolumn{3}{c}{$E$}  \\
\hline 
a&0&a  \\ 
$x_1$&0&b  \\ 
$x_2$&0&$x_1$  \\ 
\multicolumn{3}{c}{$\ldots$}   \\ 
$x_n$&0&$x_{n-1}$  \\ 
\multicolumn{3}{c}{$\ldots$}   
\end{tabular}
\end{minipage}
\begin{minipage}[b]{0.50\linewidth}
\centering
\begin{tabular}{ll} 
\multicolumn{2}{c}{$L$}  \\
\hline 
a&b  \\ 
a&$x_1$  \\ 
a&$x_2$  \\ 
\multicolumn{2}{c}{$\ldots$}  \\
a&$x_n$  \\ 
\multicolumn{2}{c}{$\ldots$}  
\end{tabular}
\end{minipage}

On the other hand, it is clear that the reduction system
in uniformly terminating.

The undecidability result can still be obtained for the $\sctaa$ class
if we allow denial constraints.
Then we simply define
$\Sigma^{\bot}_{\Theta} =
\{E^*(x,x) \rightarrow \bot\}\cup(\Sigma_{\Theta}\setminus S)$.

\begin{theorem}\label{ATHEO:sctaaUndecidable}
Let $\Sigma$ be a set of tgd's and one denial constraint.
The 
the membership problem $\Sigma\in \sctaa$ is {\sf coRE}-complete.
\end{theorem}

{\em Proof:} (Sketch)
Similarly to the proof of Theorem \ref{ATHEO:sctaeUndecidable}
it is easy to see that if an arbitrary instance $I$ contains 
a cycle, then the standard chase on $I$ with $\Sigma^{\bot}_{\Theta}$
will terminate on all branches.
This is because the fairness conditions
guarantees that the denial constraint will be fired,
and the chase will terminate.
$_{\blacksquare}$

\medskip

Finally, we note that using the same 
$\Sigma^{\bot}_{\Theta}$ reduction can be shown that
the classes $\soctaa$ and $\octaa$ are also {\sf coRE}-complete.

\section*{5. Guaranteed Termination}

The following Hasse diagram
summarizes
the stratification based classes and
their termination properties.

\bigskip
\begin{figure}[!htbp]
  \begin{center} 
\includegraphics[scale=0.4]{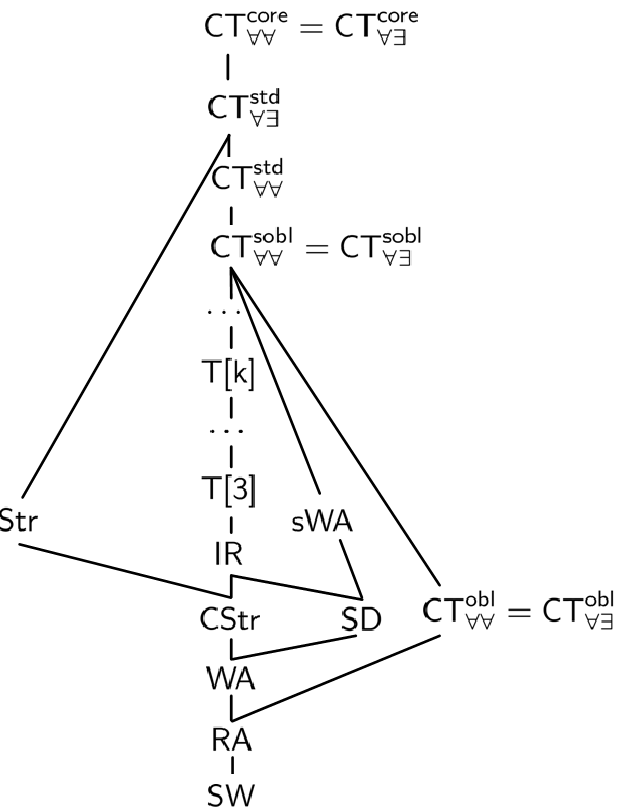}
 \caption{Sufficient classes.}
  \end{center}
\end{figure} 
\bigskip

\end{document}